\documentclass[10pt,journal,compsoc]{IEEEtran}
\IEEEoverridecommandlockouts
\usepackage{cite}
\usepackage{amsmath,amssymb,amsfonts}
\usepackage{algorithmic}
\usepackage{algorithm}
\usepackage{graphicx}
\usepackage{textcomp}
\usepackage{xcolor}
\usepackage{tikz}
\usetikzlibrary{trees, shapes, arrows.meta, calc,positioning}
\usepackage{listings}
\usepackage[hidelinks]{hyperref}
\usepackage{orcidlink}
\usepackage{newtxtext}
\lstdefinelanguage{MiniZinc}{
  keywords={include, int, var, bool, enum, array, constraint, solve, satisfy, output, forall, in, where, min, show},
  keywordstyle=\color{blue}\bfseries,
  comment=[l]{\%},
  commentstyle=\color{teal}\itshape,
  stringstyle=\color{purple},
  sensitive=true
}
\begin{document}


 \title{MASAR: Multi-stage Attack Surface Analysis and Reasoning Framework for Telecom Networks}

\author{%
\parbox{\textwidth}{\centering


{\rmfamily
\fontsize{13}{12}\selectfont
\mbox{%
Heba Osama\textsuperscript{1}\,\orcidlink{0000-0003-4550-9356}%
\hspace{0.8em}%
Wessam DeifAllah\textsuperscript{1}%
\hspace{0.8em}%
Ahmed Saafan\textsuperscript{1}\,\orcidlink{0009-0000-7927-2292}%
\hspace{0.8em}%
Haitham Ghalwash\textsuperscript{2}\,\orcidlink{0000-0001-8887-2716}%
}
\par}

\vspace{8pt}

{\footnotesize
\textsuperscript{1}Cyshield Company,
Cairo, Egypt\\[3pt]
\textsuperscript{2}Ethical Hacking and Cybersecurity,
Coventry University -- Egypt Branch,\\
hosted at The Knowledge Hub Universities,
New Cairo, Egypt
\par}

\vspace{8pt}

{\scriptsize
\begin{tabular}{@{}ll@{}}
\href{mailto:heba.osama@cyshield.com}
{heba.osama@cyshield.com}
&
\href{mailto:wessam.deifallah@cyshield.com}
{wessam.deifallah@cyshield.com}
\\[3pt]
\href{mailto:ahmed.saafan@cyshield.com}
{ahmed.saafan@cyshield.com}
&
\href{mailto:haitham.ghalwash@tkh.edu.eg}
{haitham.ghalwash@tkh.edu.eg}
\end{tabular}
\par}

}%
}

\maketitle

\begin{abstract}
The increasing complexity of modern telecommunications networks introduces security challenges that require systematic threat analysis. However, a gap remains between architectural threat modeling frameworks, such as Spoofing, Tampering, Repudiation, Information Disclosure, Denial of Service, and Elevation of Privilege (STRIDE), and operational knowledge bases, such as the MITRE Five-G Hierarchy of Threats (FiGHT). This gap limits the ability to connect architectural weaknesses with specific protocol behaviors and attack sequences.

 This paper presents Multi-stage Attack Surface Analysis and Reasoning (MASAR), a formal framework that bridges this gap through three layers: strategic threat mapping, logical attack path decomposition using directed acyclic graphs (DAGs), and operational synthesis of tactics, techniques, and procedures (TTPs). 
MASAR is evaluated using a 5G-ready IP Multimedia Subsystem (IMS) architecture, examining two signaling attack scenarios: early media billing bypass through Session Initiation Protocol (SIP) signaling and subscriber information disclosure and topology discovery through the Diameter Cx interface.
MiniZinc constraint models are used to analyze the reachability of unsafe protocol states. The analysis demonstrates how MASAR connects architectural threats with specific protocol behaviors and supports security controls that account for network states in telecommunications systems.

\end{abstract}

\begin{IEEEkeywords}
Threat Modeling Methodology, Telecom Cybersecurity, Cross-Layer Synthesis, MITRE FiGHT, STRIDE, TTP Formalization.
\end{IEEEkeywords}

\section{Introduction}
\label{sec:introduction}

The evolution of telecommunications networks toward cloud-native architectures has introduced new security challenges. Fifth Generation (5G) networks rely on a Service-Based Architecture (SBA), where network functions communicate through standardized interfaces and Application Programming Interfaces (APIs), as defined by the Third Generation Partnership Project (3GPP) in TS 23.501 \cite{3gpp23501}. Although these technologies improve network flexibility and service deployment, they also introduce additional interfaces, dependencies, and potential attack paths. Recent studies have highlighted security threats affecting different 5G network domains and the challenges associated with their increasingly complex architectures \cite{gong2025survey,tariq2025survey}.

The security architecture defined in 3GPP TS 33.501 provides mechanisms for protecting network functions and their communication interfaces \cite{3gpp33501}. However, assessing individual security mechanisms does not necessarily reveal how weaknesses across multiple interfaces can be combined into a complete attack sequence. Recent research has emphasized the importance of analyzing interconnected vulnerabilities and multistage attacks in 5G environments \cite{breeden2026survey}. This creates a need for threat analysis methods that consider both architectural weaknesses and the protocol behaviors through which attacks may occur.

Existing security approaches address these challenges from different perspectives. Architectural threat modeling methods, such as Spoofing, Tampering, Repudiation, Information Disclosure, Denial of Service, and Elevation of Privilege (STRIDE) \cite{shostack2014threat}, and the Process for Attack Simulation and Threat Analysis (PASTA) \cite{ucedavelez2015risk}, help identify potential threats and weaknesses in system design. However, translating these findings into specific protocol actions and state transitions often requires additional technical analysis. In contrast, operational knowledge bases, including the MITRE Five-G Hierarchy of Threats (FiGHT) \cite{mitrefight2023}, organize known adversarial behaviors into tactics and techniques. Although both approaches are valuable, they serve different purposes and do not independently provide a complete process for tracing architectural threats to executable protocol sequences.

 This distinction motivates the development of Multi-stage Attack Surface Analysis and Reasoning (MASAR), a formal threat modeling framework that connects architectural threat identification with detailed protocol analysis. MASAR follows three layers: strategic threat mapping, logical attack path decomposition, and operational synthesis of tactics, techniques, and procedures (TTPs). Directed acyclic graphs (DAGs) are used to represent attack dependencies, while constraint-based models examine the reachability of unsafe protocol states.

Traditional threat modeling approaches provide useful methods for identifying security concerns, but their application to complex telecommunications systems often requires substantial analyst interpretation. A single communication session may involve several network functions, signaling protocols, and state transitions. Consequently, an architectural threat may remain insufficiently understood unless the conditions and sequence of actions required for exploitation are also examined.

Formal modeling provides a structured way to represent these relationships. Graph-based approaches support the analysis of attack dependencies and potential paths through interconnected system components \cite{unger2026attackgraphs}. In MASAR, formal threat analysis provides three main benefits:

\begin{itemize}
    \item \textbf{Structured Threat Decomposition:}
    Architectural threats are represented through logical attack paths, allowing their underlying conditions and technical actions to be examined systematically.

    \item \textbf{Analysis of Protocol States and Dependencies:}
    Formal models represent relationships between protocol actions and system states, supporting the identification of potential attack sequences that may not be apparent from isolated security checks.

    \item \textbf{Support for Security Verification:}
    Constraint-based analysis enables the examination of security properties and the identification of reachable unsafe states under explicitly defined assumptions.
\end{itemize}

The MASAR framework applies these principles to a 5G-ready Internet Protocol (IP) Multimedia Subsystem (IMS) architecture.
The evaluation examines two signaling attack scenarios:
early-media billing bypass involving the Session
Initiation Protocol (SIP), and unauthorized subscriber
and routing information disclosure through the
Diameter Cx interface.
These scenarios illustrate how architectural trust assumptions and protocol interactions can be analyzed to identify potential security weaknesses.

The primary contributions of this paper are summarized as follows:

\begin{itemize}
    \item \textbf{MASAR Threat Modeling Framework:}
    A three-layer methodology connecting architectural threat identification, logical attack path decomposition, and operational threat analysis.

    \item \textbf{Formal Attack Path Analysis:}
    A graph-based representation of attack dependencies, combined with constraint models for examining protocol states and potential violations of security properties.

    \item \textbf{IMS Security Evaluation:}
    An application of MASAR to two signaling attack scenarios, illustrating the relationship between architectural weaknesses, protocol behaviors, and potential security controls.
\end{itemize}

The remainder of this paper is organized as follows.
Section~\ref{sec:related_work} reviews existing research on architectural threat modeling, telecom security knowledge bases, attack graph analysis, and protocol security.
Section~\ref{sec:methodology} presents the MASAR methodology, including strategic threat mapping, logical attack path decomposition, and operational TTP synthesis.

Section~\ref{sec:evaluation} evaluates MASAR
using two IMS signaling attack scenarios:
early-media billing bypass through SIP and
unauthorized information disclosure through Diameter Cx.

Section~\ref{sec:formal_verification} presents
the constraint-based models, security properties,
and verification results demonstrating unsafe-state
reachability under the specified model assumptions.

Section~\ref{sec:discussion} discusses the findings, limitations, and proposed architectural countermeasures.
Finally, Section~\ref{sec:conclusion} concludes the paper and outlines directions for future research.

\section{Related Work}
\label{sec:related_work}

Research relevant to MASAR spans architectural threat
modeling, telecommunications security knowledge bases,
graph-based attack analysis, and empirical protocol
security research. These areas provide complementary
perspectives, ranging from identifying architectural
weaknesses and classifying adversarial behavior to
analyzing attack dependencies and protocol-level
vulnerabilities.

Table~\ref{tab:framework_comparison} categorizes the
relevant methodologies, knowledge bases, security
guidance, and modeling techniques according to their
primary purposes and relationships to MASAR.

Building on these complementary approaches, MASAR
integrates architectural threat identification, logical
attack path decomposition, and operational TTP synthesis
within a structured three-layer methodology. Constraint-
based modeling further supports the examination of
protocol states and unsafe-state reachability under
explicitly defined assumptions.

\begin{table*}[h]
\centering
\caption{Relevant Security Approaches and Their Relationship to MASAR}
\label{tab:framework_comparison}
\footnotesize
\renewcommand{\arraystretch}{1.2}
\setlength{\tabcolsep}{4pt}

\begin{tabular}{|p{2.8cm}|p{3.5cm}|p{4.0cm}|p{5.2cm}|}
\hline
\textbf{Approach / Resource} &
\textbf{Category} &
\textbf{Primary Purpose} &
\textbf{Relationship to MASAR} \\
\hline

STRIDE \cite{shostack2014threat} &
Threat modeling methodology &
Architectural threat identification and classification &
Used in Layer 1 to identify and classify architectural threats. \\
\hline

MITRE FiGHT \cite{mitrefight2023} &
Telecom threat knowledge base &
Classification of adversarial tactics and techniques &
Used in Layer 3 to map protocol actions to existing or candidate techniques. \\
\hline

Bhadra \cite{rao2023bhadra} &
Telecom threat taxonomy &
Classification of threats affecting mobile networks &
Provides related background for telecom-specific threat categorization. \\
\hline

GSMA FS.31 \cite{gsmaFS31} &
Telecom security guidance &
Baseline security controls for mobile network operators &
Provides general security control guidance relevant to telecom network infrastructure. \\

\hline

Attack Graph 

Approaches
\cite{palma2025progressive,sainthilaire2025attackdefense} &
Graph-based security analysis &
Representation of attack dependencies and relationships &
Provide related foundations for the DAG-based decomposition used in Layer 2. \\
\hline

MiniZinc \cite{minizinc2024} &
Constraint modeling language &
Formal representation and analysis of security constraints &
Used to model protocol states and examine unsafe-state reachability. \\
\hline

\end{tabular}
\end{table*}

\subsection{Architectural Threat Modeling Frameworks}

Architectural threat modeling supports the systematic identification of security weaknesses in system components, interfaces, and trust boundaries. Farooqui et al. \cite{layered2022electronics} present a layered threat model for 5G systems that maps threats to architectural layers, affected technologies, and potential entry points. STRIDE \cite{shostack2014threat} provides a structured classification of security threats, while PASTA \cite{ucedavelez2015risk} incorporates business objectives, technical risks, and attack scenarios. Although these methods provide valuable foundations for security assessment, translating architectural findings into specific signaling procedures and protocol state transitions requires further analysis.

Recent research has examined security threats across increasingly complex telecommunications architectures. Tariq et al. \cite{tariq2025survey} investigate security threats in private 5G and beyond 5G networks through layered analysis. Breeden et al. \cite{breeden2026survey} examine security vulnerabilities in 5G network slicing and the relationships between potential attack scenarios. These studies highlight the importance of considering dependencies across network components rather than treating security threats as isolated events.

Graph-based approaches provide structured representations of attack dependencies. Kordy et al. \cite{kordy2014dag} examine foundations for attack and defense modeling, while Wide\l{} et al. \cite{widel2019beyond} survey formal methods for analyzing and generating attack trees. Konsta et al. \cite{konsta2024survey} review approaches for automatically generating attack trees and attack graphs. More recently, Palma et al. \cite{palma2025progressive} propose a scalable approach to attack graph generation that supports progressive analysis of potential attack paths. Zhang et al. \cite{zhang2025attackgraphs} investigate automated attack graph generation using graph neural networks and Bayesian models for industrial networks. Similarly, Saint-Hilaire et al. \cite{sainthilaire2025attackdefense} present an approach for generating attack and defense graphs to support the analysis of security events and potential countermeasures.

\subsection{Telecom Knowledge Bases and Matrices}

Telecommunications security knowledge bases provide structured representations of adversarial behavior and support threat identification, classification, and security operations. MITRE FiGHT \cite{mitrefight2023} organizes tactics and techniques relevant to 5G networks. Similarly, the Bhadra framework \cite{rao2023bhadra} provides a taxonomy of threats affecting mobile communication systems. 
The GSM Association (GSMA) FS.31 document
\cite{gsmaFS31} complements these resources
by providing baseline security controls
for mobile network operators.

Recent research has extended telecommunications security analysis beyond threat classification. Gong et al. \cite{gong2025survey} systematically examine attacks affecting standalone 5G networks, including user equipment, radio access networks, and core network components. Jelba and Sajan \cite{jelba2026zerotrust} propose a security framework for detecting threats across the control and user planes of cloud-native 5G networks. Their approach illustrates how relationships between signaling events and network behavior can support security monitoring.

These contributions provide valuable foundations for understanding telecommunications threats and identifying adversarial behavior. However, threat classification, operational detection, and architectural threat decomposition address different stages of security analysis. MASAR focuses on connecting these stages by mapping architectural weaknesses to logical attack paths and deriving protocol actions that can support operational threat analysis.

\subsection{Empirical Protocol Security Research}

Empirical telecommunications security research has identified vulnerabilities affecting mobile communication protocols and services. Rupprecht et al. \cite{rupprecht2020revolte} demonstrate the ReVoLTE attack, which exploits keystream reuse in Voice over Long-Term Evolution (VoLTE) communications. Lee et al. \cite{lee2022vwanalyzer} introduce VWAnalyzer to systematically examine security weaknesses in Voice over Wi-Fi (VoWiFi) protocols. While initial IMS research focused on VoLTE, modern 5G Standalone (SA) deployments rely on Voice over New Radio (VoNR), where the IMS core remains the central control plane for media session establishment over 5G Core (5GC) SBI and N5/Rx interfaces.

Research involving IMS has also demonstrated weaknesses associated with signaling procedures and session management. Li et al. \cite{li2015volte} investigate vulnerabilities in operational VoLTE networks, including signaling and billing related attacks. Lu et al. \cite{lu2020ghost} examine ghost calls and demonstrate how IMS signaling behavior can be exploited to disrupt communication services. These findings highlight the importance of analyzing protocol interactions and session states when evaluating telecommunications security.

Recent research provides additional perspectives on complex attack behavior. Gong et al. \cite{gong2025survey} review attacks across standalone 5G architectures, while Saint-Hilaire et al. \cite{sainthilaire2025attackdefense} present an automated approach for generating attack and defense graphs. Their framework connects attack relationships with potential countermeasures, demonstrating the value of linking security analysis with operational response.

Although existing studies provide substantial evidence of protocol vulnerabilities and advanced attack modeling techniques, their objectives differ from the architectural transformation process proposed in MASAR. The framework combines strategic threat mapping, logical path decomposition, and operational TTP synthesis. Constraint-based analysis is then applied to examine whether selected protocol scenarios can reach unsafe states under specified assumptions.

Together, the reviewed studies establish foundations for architectural threat analysis, graph-based security modeling, telecom threat intelligence, and empirical protocol evaluation. MASAR integrates these perspectives into a methodology focused on the relationships between architectural threats, protocol behavior, and security properties.

\section{Methodology: Cross-Layer Synthesis}
\label{sec:methodology}

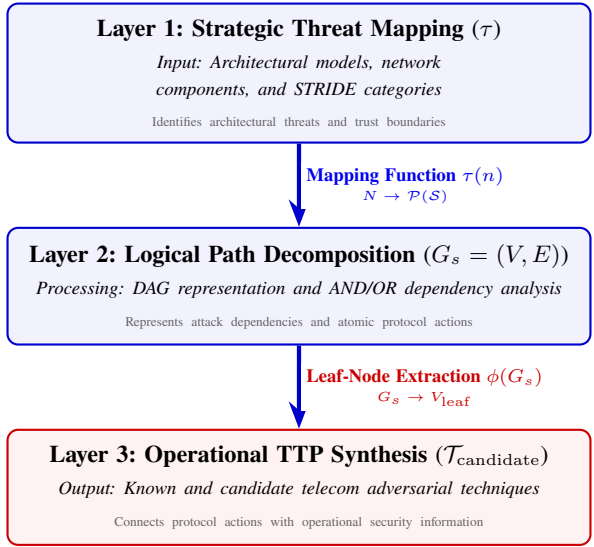
\begin{figure}[t!]
\centering
\begin{tikzpicture}[
    node distance=1.1cm and 0cm,
    auto,
    >=Stealth,
    layer/.style={
        draw=blue!80!black,
        fill=blue!5,
        thick,
        rectangle,
        rounded corners=4pt,
        align=center,
        text width=0.82\columnwidth,
        inner sep=6pt,
        font=\small
    },
    func/.style={
        font=\scriptsize\bfseries,
        text=blue!95!black,
        fill=white,
        inner sep=2pt,
        align=center
    }
]

\node[layer] (layer1) {
    \textbf{Layer 1: Strategic Threat Mapping}
    ($\tau$)\\[2pt]
    \textit{\scriptsize Input: Architectural models,
    network components, and STRIDE categories}\\[2pt]
    \textcolor{gray!80!black}{
    \tiny Identifies architectural threats and trust boundaries}
};

\node[layer, below=of layer1] (layer2) {
    \textbf{Layer 2: Logical Path Decomposition}
    ($G_s=(V,E)$)\\[2pt]
    \textit{\scriptsize Processing: DAG representation
    and AND/OR dependency analysis}\\[2pt]
    \textcolor{gray!80!black}{
    \tiny Represents attack dependencies and atomic protocol actions}
};

\node[
    layer,
    below=of layer2,
    fill=red!5,
    draw=red!80!black
] (layer3) {
    \textbf{Layer 3: Operational TTP Synthesis}
    ($\mathcal{T}_{\mathrm{candidate}}$)\\[2pt]
    \textit{\scriptsize Output: Known and candidate
    telecom adversarial techniques}\\[2pt]
    \textcolor{gray!80!black}{
    \tiny Connects protocol actions with operational security information}
};

\draw[->, ultra thick, blue!80!black]
(layer1) -- node[func] {
    Mapping Function $\tau(n)$\\[-1pt]
    \tiny $N \rightarrow \mathcal{P}(\mathcal{S})$
} (layer2);

\draw[->, ultra thick, blue!80!black]
(layer2) -- node[func, text=red!80!black] {
    Leaf-Node Extraction $\phi(G_s)$\\[-1pt]
    \tiny $G_s \rightarrow V_{\mathrm{leaf}}$
} (layer3);

\end{tikzpicture}

 \caption{The MASAR workflow from architectural threat mapping ($\tau$) to attack path decomposition ($\phi$) and TTP synthesis ($g_{\mathrm{TTP}}$).}
\label{fig:workflow}
\end{figure}

The MASAR framework provides a structured methodology for connecting architectural security threats with specific protocol behaviors and operational threat information. As illustrated in Fig.~\ref{fig:workflow}, the methodology consists of three layers: (1) Strategic Threat Mapping, (2) Logical Path Decomposition, and (3) Operational TTP Synthesis. Each layer builds on the output of the previous layer to establish a traceable relationship between architectural weaknesses, potential attack paths, and adversarial techniques.

\subsection{Layer 1: Strategic Threat Mapping}

The first layer identifies potential security threats by examining the functional architecture, network components, interfaces, and trust boundaries. STRIDE \cite{shostack2014threat} is used to classify threats associated with each component and its interactions with other network functions.

Let $N=\{n_1,n_2,\ldots,n_k\}$ represent the set of network components, and let $\mathcal{S}$ denote the set of STRIDE threat categories. The strategic mapping function is defined as (Eq.\ref{eq:strategic_mapping}):

\begin{equation}
\label{eq:strategic_mapping}
\tau: N \rightarrow \mathcal{P}(\mathcal{S}),
\end{equation}

where $\mathcal{P}(\mathcal{S})$ represents the power set of the STRIDE categories. For each component $n_i \in N$, the function $\tau(n_i)$ identifies the set of potential threats associated with that component.

This mapping provides a structured basis for examining architectural weaknesses, such as insufficient validation of signaling messages, implicit trust between network functions, or inadequate protection of communication interfaces. Identified threats are associated with the relevant components and trust boundaries to maintain traceability throughout the subsequent analysis.

However, identifying a threat category does not establish how the threat can be exploited. For example, classifying a component under information disclosure does not reveal the signaling sequence or protocol conditions that could expose sensitive information. Therefore, the threats identified in Layer 1 are examined further through logical attack path decomposition.

\subsection{Layer 2: Logical Path Decomposition}

The second layer translates architectural threats into structured attack paths by examining the conditions and protocol actions required to reach an attacker objective. DAGs are used to represent dependencies between attack steps, while AND and OR gates describe their logical relationships. This representation builds on established approaches to attack and defense modeling \cite{kordy2014dag,konsta2024survey}.

MASAR represents attack dependencies using directed
acyclic graphs (DAGs), with attack trees treated as
a restricted hierarchical form of this representation.
While attack trees organize prerequisites into
distinct branches, the general DAG representation
also supports shared dependencies across multiple
attack paths. Consequently, the logical evaluation
rules adopted by MASAR apply to both representations.

Recent research has further demonstrated the value of formal and logical attack graph analysis. Catta et al. \cite{catta2025formal} present a formal approach for examining attack paths that may lead to prohibited system states. Their work uses game theoretic modeling and automata based verification to reason about system vulnerability. Similarly, Tayouri et al. \cite{coral2025} introduce CORAL, a framework that uses logical attack graphs to assess security risks in dynamic container environments. These studies illustrate how graph representations can support structured security analysis across complex system components.

For each identified threat $s \in \tau(n_i)$, the relevant architectural conditions and protocol interactions are examined to construct an attack graph. The attacker objective is represented by the root node, while prerequisite conditions and protocol actions are represented by the remaining nodes. Dependencies are established according to the conditions required to satisfy each attack objective.

The resulting attack graph is represented as (Eq. \ref{eq:attack_graph}):

\begin{equation}
\label{eq:attack_graph}
G_s=(V,E),
\end{equation}

where $V$ is the finite set of nodes representing attack actions or logical conditions, and $E \subseteq V \times V$ is the set of directed dependency edges.

Each directed edge $(u,v)\in E$ indicates that node $u$ contributes to satisfying the condition represented by node $v$. The prerequisite nodes of $v$ are therefore defined as (Eq. \ref{eq:children}):

\begin{equation}
\label{eq:children}
\mathrm{ch}(v)=
\{u\in V \mid (u,v)\in E\}.
\end{equation}

Each node is assigned a Boolean state $x(v)\in\{0,1\}$, where $x(v)=1$ indicates that its corresponding attack condition is satisfied. Leaf nodes represent atomic protocol actions or conditions, while internal nodes combine their dependencies through AND or OR logic.

The activation state of each node is evaluated as follows (Eq. \ref{eq:attack_activation}):

\begin{equation}
\label{eq:attack_activation}
x(v)=
\begin{cases}
a(v),
& v\in V_{\mathrm{leaf}},\\[4pt]
\displaystyle\prod_{u\in\mathrm{ch}(v)}x(u),
& \mathrm{type}(v)=\mathrm{AND},\\[8pt]
\displaystyle\max_{u\in\mathrm{ch}(v)}x(u),
& \mathrm{type}(v)=\mathrm{OR},
\end{cases}
\end{equation}

where $a(v)\in\{0,1\}$ indicates whether the atomic action or condition represented by a leaf node is satisfied. Each internal node is assigned a logical type from $\{\mathrm{AND},\mathrm{OR}\}$.

Under AND logic, all prerequisite conditions must be satisfied. Under OR logic, satisfying at least one prerequisite condition is sufficient to activate the corresponding node.

The set of leaf nodes is defined as shown (Eq.\ref{eq:leaf_nodes}):

\begin{equation}
\label{eq:leaf_nodes}
V_{\mathrm{leaf}}=
\{v\in V\mid \mathrm{ch}(v)=\emptyset\}.
\end{equation}

The leaf-node extraction function is expressed as shown in Eq. \ref{eq:leaf_extraction}:

\begin{equation}
\label{eq:leaf_extraction}
\phi(G_s)=V_{\mathrm{leaf}}.
\end{equation}

This function identifies the atomic actions and conditions represented within the attack graph. Their logical dependencies determine whether the root attack objective can be satisfied.

For a particular assignment of atomic conditions, the attack objective is satisfied when (\ref{eq:root_activation}) holds:

\begin{equation}
\label{eq:root_activation}
x(\mathrm{root}(G_s))=1.
\end{equation}

The logical representation supports the systematic examination of attack dependencies and identifies conditions relevant to successful exploitation. It also provides a basis for determining where security controls may interrupt an attack sequence.

Although the DAG represents logical dependencies, the temporal ordering of signaling messages requires additional protocol constraints. Such constraints are examined through the formal verification models presented in Section~\ref{sec:formal_verification}.

\subsection{Layer 3: Operational TTP Synthesis}

The third layer translates the atomic protocol actions identified during logical decomposition into operational threat information. Its objective is to connect technical behavior represented in the attack graph with relevant adversarial tactics and techniques.

Each identified action is examined in relation to existing telecom threat knowledge bases, particularly MITRE FiGHT \cite{mitrefight2023}. This analysis considers the associated protocol, attacker objective, required conditions, and relationship to existing techniques.

The operational mapping function is represented in (Eq. \ref{eq:ttp_mapping})

\begin{equation}
\label{eq:ttp_mapping}
g_{\mathrm{TTP}}:
V_{\mathrm{leaf}}
\rightarrow
\mathcal{T}_{\mathrm{FiGHT}}
\cup
\mathcal{T}_{\mathrm{candidate}},
\end{equation}

where $\mathcal{T}_{\mathrm{FiGHT}}$ represents existing techniques in MITRE FiGHT, and $\mathcal{T}_{\mathrm{candidate}}$ represents candidate techniques that require further assessment.

\begin{figure}[!htbp]
\centering
\resizebox{\columnwidth}{!}{%
\begin{tikzpicture}[
  box/.style={draw=blue!80!black, line width=1.2pt, rounded corners=6pt,
              fill=blue!7, align=center, minimum width=3.2cm,
              minimum height=2.2cm, inner sep=4pt},
  arr/.style={-{Stealth[length=3mm,width=2.6mm]}, line width=1.4pt,
              draw=blue!80!black},
  lbl/.style={font=\bfseries\scriptsize, text=blue!80!black,
              midway, above=2pt}
]

\node[box] (a) {%
  {\small\textbf{STRIDE Category}}\\[2pt]
  {\footnotesize $s \in \tau(n_i)$}\\[2pt]
  {\scriptsize (e.g., Info Disclosure)}};

\node[box, right=1.8cm of a] (b) {%
  {\small\textbf{DAG Attack Graph}}\\[2pt]
  {\footnotesize $G_s = (V,E)$}\\[2pt]
  {\scriptsize Analyst-guided}\\
  {\scriptsize protocol decomposition}};

\node[box, right=1.8cm of b] (c) {%
  {\small\textbf{MiniZinc CSP Model}}\\[2pt]
  {\footnotesize $\neg P_{\mathtt{safe}}$}\\
  {\footnotesize (Violation Query)}\\[2pt]
  {\scriptsize Attack conditions}\\
  {\scriptsize $\rightarrow$ CSP constraints}};

\draw[arr] (a) -- node[lbl]{Decompose} (b);
\draw[arr] (b) -- node[lbl]{Formalize} (c);

\end{tikzpicture}%
}
\caption{MASAR pipeline: from STRIDE category to MiniZinc CSP model}
\label{fig:masar_pipeline}
\end{figure}
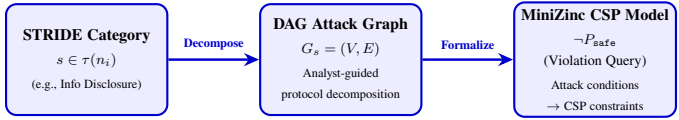
The transformation pipeline shown in Fig.~\ref{fig:masar_pipeline} connects architectural threat identification with logical
attack decomposition and constraint-based
verification. First, a STRIDE threat
$s \in \tau(n_i)$ is associated with an attacker
objective represented by the root node of an
attack DAG $G_s$. The objective is decomposed
into prerequisite conditions and protocol actions
through analysis of the relevant network
architecture, trust boundaries, and applicable
protocol specifications. Mandatory prerequisites
are represented through AND dependencies,
whereas alternative conditions are represented
through OR dependencies.

Second, the attack graph provides a structured
basis for identifying the protocol states,
conditions, and events relevant to formal
verification. These elements are represented
using finite-domain variables and constraints
in MiniZinc. Logical dependencies are modeled
through Boolean constraints, while temporal
ordering and state transitions are specified
separately according to the relevant protocol
behavior. The resulting constraint satisfaction
problem is evaluated against a defined safety
property $P_{\mathrm{safe}}$. To identify a
potential unsafe execution, the solver searches
for a satisfying assignment to
$\neg P_{\mathrm{safe}}$ under the specified
model assumptions and execution bounds.

In the current implementation, the transformation
from STRIDE threat categories to attack DAGs and
the corresponding MiniZinc constraint models is
performed manually by security researchers.
The process follows a structured, analyst-guided
methodology based on architectural analysis,
protocol specifications, and identified attack
dependencies. Automated generation of attack DAGs
and executable constraint models is outside the
scope of the present implementation and is
identified as a direction for future work.

When an identified protocol action corresponds to an existing technique, it is associated with the relevant knowledge base entry. Otherwise, the action may be documented as a candidate technique, together with its protocol context, required conditions, and potential security implications.

This distinction separates the identification of potentially unclassified behavior from the confirmation of a novel attack technique. Establishing novelty requires further comparison with existing research, documented vulnerabilities, and relevant threat knowledge bases.

The synthesis process establishes a traceable relationship between architectural threats and operational security information. Its outputs can support the development of detection requirements, security analysis procedures, and architectural countermeasures.

Table~\ref{tab:math_summary} summarizes the mathematical notation and its role in the MASAR methodology.

\begin{table}[!htbp]
\centering
\caption{Mathematical Notation Used in MASAR}
\label{tab:math_summary}
\footnotesize
\renewcommand{\arraystretch}{1.15}
\setlength{\abovecaptionskip}{4pt}
\setlength{\belowcaptionskip}{0pt}

\resizebox{\columnwidth}{!}{%
\begin{tabular}{|l|p{3.3cm}|p{3.7cm}|}
\hline
\textbf{Symbol} &
\textbf{Definition} &
\textbf{Role} \\
\hline
$N$ &
Set of network components &
Architectural elements \\
\hline
$\mathcal{S}$ &
Set of STRIDE categories &
Threat classification \\
\hline
$\tau$ &
$N \rightarrow \mathcal{P}(\mathcal{S})$ &
Strategic threat mapping \\
\hline
$G_s=(V,E)$ &
Attack graph for threat $s$ &
Logical attack representation \\
\hline
$\mathrm{ch}(v)$ &
Prerequisite nodes of $v$ &
Attack dependencies \\
\hline
$x(v)$ &
Boolean node state &
Condition satisfaction \\
\hline
$a(v)$ &
Atomic condition state &
Leaf-node evaluation \\
\hline
$V_{\mathrm{leaf}}$ &
Nodes without prerequisites &
Atomic actions and conditions \\
\hline
$\phi(G_s)$ &
Leaf-node extraction &
Logical decomposition \\
\hline
$g_{\mathrm{TTP}}$ &
Operational technique mapping &
TTP synthesis \\
\hline
$\mathcal{T}_{\mathrm{candidate}}$ &
Candidate techniques &
Further security assessment \\
\hline
\end{tabular}%
}
\end{table}
\vspace{-4pt}
\subsection{Algorithmic Representation of the Transformation Process}

The MASAR transformation process combines logical attack
graph analysis with operational technique mapping.
Algorithm~\ref{alg:masar_pipeline} provides a procedural
representation of the methodology, including the extraction
of atomic actions, evaluation of attack dependencies,
and identification of candidate operational techniques.

The algorithm describes the analytical steps followed
by security researchers rather than an implemented
automated transformation system. In the current
evaluation, STRIDE threat mapping, attack DAG
construction, and MiniZinc constraint model development
are performed manually. Automating these stages is
identified as a direction for future work.

For each attack graph, the algorithm first identifies the leaf nodes representing atomic protocol actions or conditions. It then evaluates the remaining nodes according to their logical dependencies using a topological ordering of the DAG. This ensures that prerequisite conditions are evaluated before the nodes that depend on them, following established graph traversal principles \cite{clrs2022introduction}.

The resulting atomic actions are examined against existing technique definitions to determine whether they correspond to known adversarial behavior or require further assessment.

\begin{algorithm}[htbp]
\caption{MASAR Transformation Pipeline}
\label{alg:masar_pipeline}

\begin{algorithmic}[1]

\REQUIRE Attack DAG $G_s=(V,E)$,
leaf condition values $a(v)$,
existing technique set $\mathcal{T}_{\mathrm{FiGHT}}$

\ENSURE Root condition state,
atomic actions,
candidate techniques

\STATE Identify $V_{\mathrm{leaf}}$
using Eq.~\eqref{eq:leaf_nodes}

\STATE Initialize
$\mathcal{T}_{\mathrm{candidate}}
\leftarrow \emptyset$

\STATE Obtain a topological ordering
$(v_1,\ldots,v_{|V|})$ of $G_s$

\FORALL{$v\in V_{\mathrm{leaf}}$}
    \STATE $x(v)\leftarrow a(v)$
\ENDFOR

\STATE \textit{Evaluate logical attack dependencies}

\FOR{$i=1$ to $|V|$}

    \STATE $v\leftarrow v_i$

    \IF{$v\notin V_{\mathrm{leaf}}$}

        \IF{$\mathrm{type}(v)=\mathrm{AND}$}
            \STATE
            $x(v)\leftarrow
            \prod_{u\in\mathrm{ch}(v)}x(u)$
        \ELSE
            \STATE
            $x(v)\leftarrow
            \max_{u\in\mathrm{ch}(v)}x(u)$
        \ENDIF

    \ENDIF

\ENDFOR

\STATE \textit{Map atomic actions to techniques}

\FORALL{$v\in V_{\mathrm{leaf}}$}

    \STATE Determine technique mapping
    $g_{\mathrm{TTP}}(v)$

    \IF{$g_{\mathrm{TTP}}(v)
    \notin \mathcal{T}_{\mathrm{FiGHT}}$}

        \STATE
        $\mathcal{T}_{\mathrm{candidate}}
        \leftarrow
        \mathcal{T}_{\mathrm{candidate}}
        \cup\{g_{\mathrm{TTP}}(v)\}$

    \ENDIF

\ENDFOR

\STATE \textbf{return}
$x(\mathrm{root}(G_s))$,
$V_{\mathrm{leaf}}$,
$\mathcal{T}_{\mathrm{candidate}}$

\end{algorithmic}
\end{algorithm}

The algorithm provides a structured representation of the relationship between logical attack dependencies and operational threat information. It does not independently establish protocol exploitability or the novelty of identified techniques. These properties require further validation through protocol constraints, formal verification, and comparison with existing security research.

\section{Methodology Evaluation: The IMS Core Use Case}
\label{sec:evaluation}

The MASAR methodology is evaluated using a 5G-ready
IP Multimedia Subsystem (IMS) architecture to examine
how architectural threats can be translated into
protocol-level attack conditions and operational
threat information. IMS provides a suitable evaluation
context because session establishment, subscriber
authorization, media handling, and charging involve
interactions among multiple network functions and
signaling protocols.

The evaluation considers two scenarios: Early Media
Free Ride and Unauthorized Diameter Cx Information
Disclosure. Although these scenarios target different
security objectives, both illustrate how weaknesses
in trust enforcement and interactions between protocol
procedures may create opportunities for abuse.

For each scenario, the evaluation follows the MASAR
analytical layers. Layer 1 identifies the relevant
architectural threats and network components. Layer 2
decomposes the attack objective into prerequisite
conditions and protocol actions. Layer 3 interprets
the resulting behavior in relation to operational
threat information and candidate TTPs. The corresponding
protocol-state constraints are examined through the
formal models presented in
Section~\ref{sec:formal_verification}.

\subsection{IMS Architecture, Trust Boundaries,
and End-to-End Threat Traceability}
\label{subsec:ims_traceability}

The end-to-end application of MASAR is
demonstrated through a conceptual IMS data
flow diagram (DFD), which identifies the
signaling entities, information flows,
and trust boundaries associated with
the two evaluated scenarios.

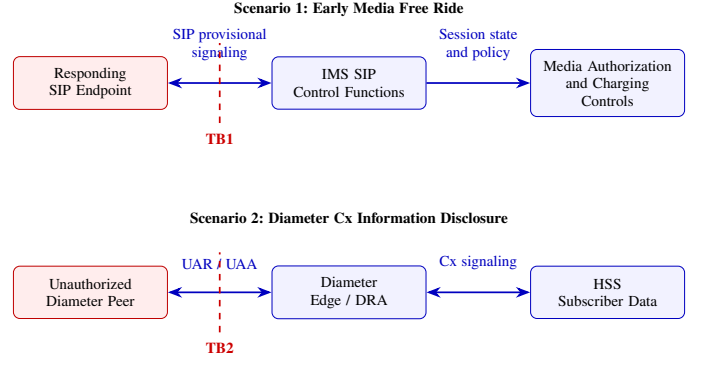
\begin{figure}[htbp]
\centering

\resizebox{\columnwidth}{!}{%
\begin{tikzpicture}[
    >=Stealth,
    font=\scriptsize,
    entity/.style={
        draw=blue!75!black,
        fill=blue!5,
        rounded corners=3pt,
        align=center,
        text width=2.15cm,
        minimum height=0.85cm,
        inner sep=5pt
    },
    external/.style={
        draw=red!75!black,
        fill=red!7,
        rounded corners=3pt,
        align=center,
        text width=2.15cm,
        minimum height=0.85cm,
        inner sep=5pt
    },
    flow/.style={
        ->,
        thick,
        blue!75!black
    }
]

\node[external] (endpoint) at (0,0)
{Responding\\SIP Endpoint};

\node[entity] (sip) at (4.2,0)
{IMS SIP\\Control Functions};

\node[entity] (media) at (8.4,0)
{Media Authorization\\and Charging Controls};

\draw[<->,thick,blue!75!black]
(endpoint) --
node[midway,above=7pt,align=center,
font=\scriptsize]
{SIP provisional\\signaling}
(sip);

\draw[flow]
(sip) --
node[midway,above=7pt,align=center,
font=\scriptsize]
{Session state\\and policy}
(media);

\draw[red!80!black,dashed,thick]
(2.1,-0.65) -- (2.1,0.65);

\node[red!80!black,font=\scriptsize\bfseries]
at (2.1,-0.90) {TB1};

\node[external] (peer) at (0,-3.4)
{Unauthorized\\Diameter Peer};

\node[entity] (edge) at (4.2,-3.4)
{Diameter\\Edge / DRA};

\node[entity] (hss) at (8.4,-3.4)
{HSS\\Subscriber Data};

\draw[<->,thick,blue!75!black]
(peer) --
node[midway,above=7pt,align=center,
font=\scriptsize]
{UAR / UAA}
(edge);

\draw[<->,thick,blue!75!black]
(edge) --
node[midway,above=7pt,align=center,
font=\scriptsize]
{Cx signaling}
(hss);

\draw[red!80!black,dashed,thick]
(2.1,-4.05) -- (2.1,-2.75);

\node[red!80!black,font=\scriptsize\bfseries]
at (2.1,-4.30) {TB2};

\node[font=\scriptsize\bfseries]
at (4.2,1.2)
{Scenario 1: Early Media Free Ride};

\node[font=\scriptsize\bfseries]
at (4.2,-2.2)
{Scenario 2: Diameter Cx Information Disclosure};

\end{tikzpicture}%
}

\caption{Conceptual IMS DFD showing the
evaluated signaling paths and trust boundaries.}
\label{fig:ims_dfd}
\end{figure}

Fig.~\ref{fig:ims_dfd} illustrates the two
signaling paths considered in the evaluation.
The first concerns SIP provisional signaling,
early-media authorization, and charging
enforcement. The second concerns Diameter Cx
authorization requests and subscriber-related
responses exchanged through the IMS signaling
infrastructure.

The DFD provides the architectural input to
MASAR Layer 1. STRIDE categories are assigned
by examining the relevant components, data flows,trust boundaries, and possible consequences of insufficient security enforcement. For the Early Media Free Ride scenario,
the relevant concern is classified under
Tampering, as manipulation of provisional
signaling behavior may exploit inadequate
coordination between session state, media
authorization, and charging controls.

For the Diameter Cx scenario, the primary
STRIDE category is Information Disclosure,
as insufficient peer authorization may
allow protected subscriber or routing
information to be returned to an
unauthorized requester. These architectural findings are then
transformed into logical attack dependencies
in Layer 2 and candidate operational
technique descriptions in Layer 3.

Table~\ref{tab:masar_traceability} summarizes
the resulting end-to-end traceability
for both scenarios.

\begin{table*}[htbp]
\centering
\caption{End-to-End MASAR Traceability
for the Evaluated IMS Attack Scenarios}
\label{tab:masar_traceability}
\footnotesize
\renewcommand{\arraystretch}{1.2}
\setlength{\tabcolsep}{4pt}

\begin{tabular}{|
p{2.7cm}|
p{2.8cm}|
p{2.5cm}|
p{3.5cm}|
p{3.3cm}|}
\hline
\textbf{Scenario} &
\textbf{DFD Element and Trust Boundary} &
\textbf{Layer 1: STRIDE} &
\textbf{Layer 2: Logical Attack Path} &
\textbf{Layer 3: Candidate TTP} \\
\hline

Early Media Free Ride &
SIP endpoint, IMS signaling control,
and media/charging authorization
across TB1 &
Tampering &
Establish provisional media and
exploit insufficient charging or
duration enforcement through
continued provisional signaling &
Provisional-State Early-Media
Exploitation \\
\hline

Unauthorized Diameter Cx
Information Disclosure &
Diameter peer, signaling edge,
and HSS access across TB2 &
Information Disclosure &
Obtain signaling access, submit
a UAR, bypass insufficient peer
authorization, and receive a UAA
containing protected information &
Unauthorized Diameter Cx
Subscriber and Topology
Information Disclosure \\
\hline

\end{tabular}
\end{table*}

The traceability table demonstrates
how the two scenarios progress from
architectural threat identification
to logical attack conditions and
operational TTP synthesis. The detailed logical decomposition
and protocol-level interpretation
of each scenario are presented in
the following subsections. The
corresponding constraint models
are provided in
Section~\ref{sec:formal_verification}.
The diagram and mappings represent
the architectural assumptions used
for this evaluation. They do not
constitute validation of a specific
deployed IMS topology.

\subsection{Attack Vector 1: Early Media Free Ride
(Zero Answer Fraud)}

This scenario examines how interactions between SIP
session establishment, early-media authorization, and
charging enforcement can create a potential opportunity
for unauthorized media consumption. The analysis first
establishes the technical assumptions and attacker
capabilities, then decomposes the attack into logical
conditions, and finally interprets the resulting
behavior from an operational threat perspective.

\subsubsection{Technical Scenario and Attack Preconditions}

The first stage establishes the protocol context and
configuration assumptions required to analyze the
attack. In particular, the scenario considers whether
media can remain authorized during a provisional SIP
session without triggering the charging control
expected under the modeled deployment configuration.

This scenario examines a potential inconsistency
between early-media handling and charging enforcement
in an IMS environment \cite{ietfEarlyMedia3960}.
SIP supports media exchange before final session
establishment, including during provisional signaling
associated with \texttt{183 Session Progress} responses.

IMS charging behavior depends on the applicable
charging architecture, network functions, and
operator-configured trigger conditions. Consequently,
a final \texttt{200 OK} response is not universally
the first or only charging trigger. In the vulnerable
configuration considered by MASAR, early media is
authorized during the provisional session state,
while the charging control under examination does
not adequately account for this media usage before
final session confirmation.

An adversary may exploit this configuration by
prolonging provisional signaling and maintaining
RTP/SRTP media transmission without triggering
the expected charging control. The potential bypass
depends on insufficient early-media duration
restrictions, inadequate coordination between
session-state and charging policies, or ineffective
enforcement of early-media authorization mechanisms
\cite{ietfEarlyMedia5009}.

The scenario begins when a SIP INVITE request
initiates a communication session, following the
signaling procedures defined in RFC 3261
\cite{ietfSIP3261}.
A responding endpoint or an intermediary may return
a \texttt{183 Session Progress} response containing
a Session Description Protocol (SDP) body. Depending
on signaling exchange and applicable session policies,
this response can contribute to the establishment
of an early media path.

The Real-time Transport Protocol (RTP) may then
carry media before the session receives a final
\texttt{200 OK} response. In the modeled vulnerable
configuration, media authorization is allowed during
the provisional state, while the charging mechanism
is assumed to depend on final session confirmation.
An adversary controlling the responding endpoint
may attempt to delay or withhold the final
\texttt{200 OK} response while maintaining the
conditions necessary for early-media transmission.
If the network lacks effective media-duration limits,
charging controls, or session-state enforcement,
media may continue to flow without being accounted
for as intended.

This scenario therefore illustrates a potential
weakness in the coordination of SIP session-state
management, media authorization, and charging
enforcement. Its feasibility must be established
against the specific IMS configuration and charging
mechanisms under evaluation; it should not be
assumed to arise from the use of early media alone.
\subsubsection{Attack Path and Logical Decomposition}

Following the technical scenario definition, the
potential attack is decomposed into a logical
structure representing the conditions required
to achieve the adversary's objective. Consistent
with MASAR Layer 2, the decomposition distinguishes
mandatory prerequisites from alternative actions
that may contribute to the same attack outcome.

\begin{figure}[htbp]
\centering
\resizebox{\columnwidth}{!}{%
\begin{tikzpicture}[
    level 1/.style={
        sibling distance=4.8cm,
        level distance=1.6cm
    },
    level 2/.style={
        sibling distance=2.4cm,
        level distance=1.6cm
    },
    every node/.style={
        draw,
        rectangle,
        rounded corners=3pt,
        align=center,
        fill=blue!5,
        font=\scriptsize,
        text width=2.1cm,
        inner sep=3pt
    },
    edge from parent/.style={
        draw,
        thick,
        Stealth-
    }
]
\node[
    fill=blue!15,
    font=\bfseries\scriptsize
] (root) {
    Goal: Execute\\Zero Answer Fraud\\(AND)
}
child {
    node (G1) {
        G1: Establish\\Early Media\\(AND)
    }
    child {
        node (G11) {
            1.1 Receive\\SIP 183 Response
        }
    }
    child {
        node (G12) {
            1.2 Negotiate\\Early SDP
        }
    }
}
child {
    node (G2) {
        G2: Avoid\\Charging Trigger\\(OR)
    }
    child {
        node[
            fill=red!15,
            draw=red!80!black
        ] (G21) {
            2.1 Suppress\\Final 200 OK
        }
    }
    child {
        node (G22) {
            2.2 Delayed 4xx\\
            Rejection After\\Media Stream
        }
    }
};
\end{tikzpicture}%
}

\caption{DAG-based attack decomposition for
the Early Media Free Ride scenario,
represented in attack-tree form.}
\label{fig:early_media_tree_v2}
\end{figure}
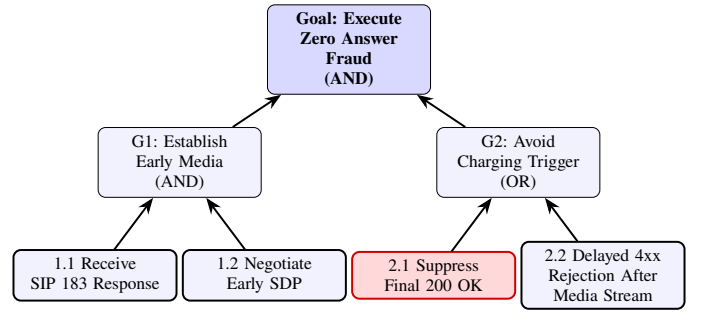

Fig.~\ref{fig:early_media_tree_v2} illustrates the
logical decomposition of the modeled attack.

Subgoal G2 represents avoiding the charging control
expected under the assumed deployment configuration.
Node 2.1 represents withholding final SIP session
confirmation, while Node 2.2 represents delayed
session rejection following provisional media
transmission. Under insufficient charging and
early-media enforcement, either behavior may
contribute to unauthorized media consumption.
A subsequent \texttt{4xx} response terminates
the attempted session but does not necessarily
eliminate the accounting obligations associated
with media already transmitted.

The resulting logical path connects provisional
session establishment with continued media
authorization and the absence of the expected
charging event. Its success depends on the
configuration assumptions identified in the
preceding subsection, particularly the lack of
effective controls limiting uncharged early media.

\subsubsection{Operational Interpretation and TTP Mapping}

The final stage interprets the identified protocol
actions in relation to operational threat behavior,
consistent with MASAR Layer 3. This step connects
the logical attack conditions with candidate
technique descriptions without assuming that
the behavior constitutes a previously unrecognized
or independently validated attack technique.

The early-media scenario involves manipulation
of provisional SIP session behavior to maintain
media transmission without the expected charging
event. The relevant actions include maintaining
provisional signaling, delaying final session
confirmation, and exploiting insufficient
coordination between media authorization and
charging controls.

These behaviors are examined in relation to
telecommunications threat knowledge bases,
including MITRE FiGHT \cite{mitrefight2023}.
The resulting behavior is documented as a
candidate technique, Provisional-State Early-Media
Exploitation, for further assessment and
classification.

The analysis also identifies potential security
control requirements, including early-media
duration restrictions, session-state-aware media
authorization, and coordination between signaling
and charging enforcement.
The corresponding protocol-state assumptions
and unsafe-state reachability condition are
formalized in
Section~\ref{sec:formal_verification}.
\subsection{Attack Vector 2: Unauthorized Diameter
Cx Information Disclosure}

The second scenario examines potential disclosure
of subscriber authorization or internal IMS
routing information through the Diameter Cx
interface. As in the first scenario, the
analysis begins with the technical preconditions,
then identifies the logical attack path, and
finally connects the observed protocol behavior
with operational threat information.

\subsubsection{Technical Scenario and Attack Preconditions}
This stage identifies the network functions,
signaling procedures, and authorization assumptions
relevant to the modeled information disclosure
scenario. The primary concern is whether a
Diameter peer lacking the required authorization
can submit subscriber-related requests and
receive information intended for trusted
IMS signaling entities.

The Cx interface supports communication between
IMS control functions and subscriber information
systems. The User-Authorization-Request (UAR)
and User-Authorization-Answer (UAA) procedures
are defined in 3GPP TS 29.229 \cite{3gpp29229}.

The modeled security concern arises when an
unauthorized Diameter peer can issue subscriber
authorization requests that are accepted by
the network and obtain response information
that reveals subscriber status or internal
routing details.

Under normal operation, the Interrogating
Call Session Control Function (I-CSCF)
exchanges Diameter messages with the Home
Subscriber Server (HSS) through the Cx
interface. The underlying Diameter protocol
specifies peer relationships and message
exchange procedures \cite{ietfDiameter6733}.

A UAR carries information required for the
subscriber authorization procedure, and the
corresponding UAA returns the result and
associated information defined by the
applicable protocol procedure.

The scenario assumes that an adversary has
obtained access to a signaling path capable
of reaching the relevant Diameter service
and can cause requests to be accepted as
originating from an authorized peer. This
access and acceptance condition is an
explicit prerequisite of the modeled
attack, not an inherent property of the
Cx interface.

The adversary generates a UAR containing
a target subscriber identity. If the
Diameter edge infrastructure accepts and
forwards the request, the HSS processes
it according to the applicable
authorization procedure.

The response may provide information that
helps distinguish subscriber authorization
outcomes. Depending on the procedure and
response content, additional routing
information may also be exposed.

Repeated queries could therefore support
subscriber presence enumeration or reveal
information about the internal IMS topology
when peer authorization and information
disclosure controls are insufficient.

This scenario highlights the distinction
between Diameter peer authorization and
SIP-level subscriber authentication.
These mechanisms operate at different
stages of IMS signaling and provide
different security functions; the
presence of one does not automatically
establish the effectiveness of the other.

Two information-disclosure opportunities
are considered:

\begin{itemize}

    \item \textbf{Subscriber Authorization Enumeration:}
    Differences between successful authorization
    outcomes and subscriber-related error responses
    may reveal whether a queried identity is
    recognized or satisfies particular authorization
    conditions.

    Relevant 3GPP experimental result codes include
    \texttt{DIAMETER\_ERROR\_USER\_UNKNOWN}
    (5001) and
    \texttt{DIAMETER\_ERROR\_IDENTITIES\_DONT\_MATCH}
    (5002).

    These responses do not independently establish
    that the subscriber is currently registered
    or active.

    \item \textbf{Internal IMS Routing Information:}
    Depending on the authorization procedure and
    applicable response conditions, the UAA may
    include a \texttt{Server-Name} AVP identifying
    an assigned S-CSCF or a
    \texttt{Server-Capabilities} AVP containing
    information relevant to S-CSCF selection.

    Disclosure of these attributes to an
    unauthorized peer may expose internal routing
    characteristics or network topology.

\end{itemize}

\subsubsection{Attack Path and Logical Decomposition}

Following the identification of the required
preconditions, the scenario is decomposed into
a sequence of dependent signaling actions.
This corresponds to MASAR Layer 2, in which
an architectural information-disclosure threat
is translated into explicit conditions and
protocol-level actions.

Layer 1 identifies the relevant information
disclosure threat and the trust boundary
between Diameter signaling peers and
subscriber information services.

Layer 2 represents the scenario through the
following sequence of required conditions:

\begin{enumerate}

    \item The adversary obtains access to a
    Diameter signaling path capable of
    reaching the relevant authorization service.

    \item The adversary sends a UAR using
    a selected subscriber identity.

    \item The signaling infrastructure
    accepts and forwards the request.

    \item The HSS processes the request
    and generates a UAA.

    \item The returned information allows
    the adversary to infer subscriber
    authorization status or, under
    additional conditions, internal
    routing information.

\end{enumerate}

These actions depend on the successful
satisfaction of the preceding conditions.
The modeled attack therefore requires both
signaling access and insufficient enforcement
of the relevant authorization boundaries.

\begin{figure}[htbp]
\centering

\resizebox{\columnwidth}{!}{%
\begin{tikzpicture}[
    node distance=1.0cm,
    font=\scriptsize,
    >=Stealth,
    box/.style={
        draw,
        fill=gray!10,
        text width=2.0cm,
        minimum height=0.65cm,
        align=center,
        rounded corners=3pt,
        font=\bfseries\scriptsize
    }
]

\node[
    box,
    fill=red!15,
    draw=red
] (Rogue) {
    Rogue Diameter\\Peer
};

\node[
    box,
    right=1.15cm of Rogue
] (DRA) {
    Diameter\\Edge / DRA
};

\node[
    box,
    right=1.15cm of DRA
] (HSS) {
    HSS
};

\draw[dashed,gray]
(Rogue.south) -- ++(0,-3.3);

\draw[dashed,gray]
(DRA.south) -- ++(0,-3.3);

\draw[dashed,gray]
(HSS.south) -- ++(0,-3.3);

\draw[->,red,thick]
([yshift=-0.65cm]Rogue.south)
--
node[
    above,
    font=\tiny\bfseries
] {1. UAR}
([yshift=-0.65cm]DRA.south);

\draw[->,red,thick]
([yshift=-1.3cm]DRA.south)
--
node[
    above,
    font=\tiny\bfseries
] {2. Forward UAR}
([yshift=-1.3cm]HSS.south);

\draw[->,blue!80!black,dashed,thick]
([yshift=-2.0cm]HSS.south)
--
node[
    above,
    font=\tiny\bfseries
] {3. UAA}
([yshift=-2.0cm]DRA.south);

\draw[->,blue!80!black,dashed,thick]
([yshift=-2.7cm]DRA.south)
--
node[
    above,
    font=\tiny\bfseries,
    align=center
] {4. Response Information}
([yshift=-2.7cm]Rogue.south);

\end{tikzpicture}%
}

\caption{UAR-based presence enumeration via the HSS.}
\label{fig:uar_sequence}
\end{figure}
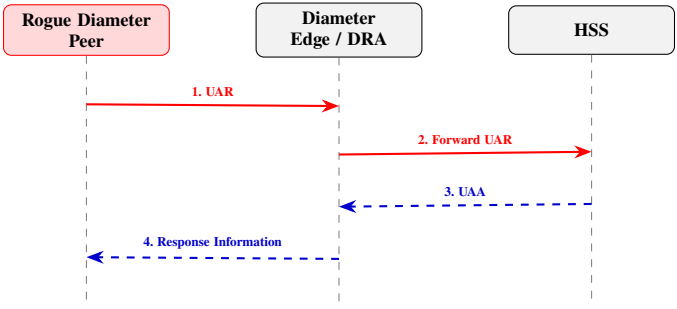

Figure~\ref{fig:uar_sequence} illustrates the
message exchange considered in this scenario.

The sequence illustrates the causal
relationships between unauthorized signaling
access, request acceptance, subscriber
authorization processing, and response
delivery. These dependencies provide the
basis for the protocol ordering constraints
used in the corresponding formal model.

\subsubsection{Operational Interpretation and TTP Mapping}

The final stage examines how the modeled
Diameter signaling behavior relates to
operational threat information. Consistent
with MASAR Layer 3, the objective is to
interpret the protocol actions and identify
potential technique descriptions rather
than assume that every modeled behavior
represents a novel or confirmed attack.

The use of unauthorized Diameter queries
to infer subscriber information or
routing characteristics may be documented
as a candidate reconnaissance technique.

More specifically, the behavior is
characterized in this study as the candidate
technique Unauthorized Diameter Cx
Subscriber and Topology Information
Disclosure, subject to further assessment
against existing telecom threat knowledge
bases, including MITRE FiGHT
\cite{mitrefight2023}.

The scenario also identifies potential
control requirements, including Diameter
peer authentication, signaling access
restrictions, request authorization, and
validation of information returned
across trust boundaries.

The corresponding protocol ordering
assumptions and security properties
are examined through the constraint
model presented in
Section~\ref{sec:formal_verification}.

\section{Formal Verification via Constraint Satisfaction}
\label{sec:formal_verification}

To examine the feasibility of the attack scenarios
identified by MASAR, constraint satisfaction is used
to analyze protocol states and their relationships.
The verification process represents relevant signaling
procedures as finite sets of variables and constraints,
allowing the reachability of potentially unsafe states
to be examined.

The analysis focuses on two scenarios introduced
in Section~\ref{sec:evaluation}: Early Media Free
Ride and unauthorized Diameter Cx information
disclosure. The first examines the relationship
between SIP session states, media authorization,
and charging. The second examines whether insufficient
Diameter peer authorization enforcement can permit
protected subscriber or routing information to be
disclosed to an unauthorized requester.

Each scenario is represented as a constraint
satisfaction problem (CSP). A security property is
defined to describe the expected safe behavior, and
its negation is used to search for a violating execution
trace. A satisfiable result indicates that a violating
state is reachable under the modeled constraints.
An unsatisfiable result indicates that no such state
exists within the specified model and search bounds.

The models are implemented using MiniZinc
\cite{minizinc2024}, a constraint modeling language
that supports finite-domain variables and logical
constraints. The corresponding executable
specifications are provided in
Appendix~\ref{app:minizinc_models}.

For consistency, both models are presented using four
analytical stages: model representation and assumptions,
protocol constraints, security properties and reachability
analysis, and witness trace and result interpretation.
Although the models address different signaling protocols,
this structure provides a common basis for describing
their formulation and interpreting their outcomes.

\subsection{Model 1: Early Media Billing Bypass}

The first model examines a potential inconsistency
between SIP session progression, media authorization,
and charging enforcement. It considers a network
configuration in which media transmission is permitted
during a provisional session state, while the charging
event under investigation is associated with final
session confirmation.

The objective is to determine whether the modeled
session can remain in a provisional state with media
transmission authorized beyond a specified duration
without the expected charging event. This analysis
evaluates the reachability of the defined unsafe
condition under the model's assumptions; it does not,
by itself, establish that the same behavior occurs
in every IMS implementation.

\subsubsection{Model Representation and Assumptions}
\label{subsec:state_encoding}

The model represents the session as a sequence of
$T$ discrete steps. Each step corresponds to a
state transition or an observation of the signaling
process.

At each step $t \in \{1,\ldots,T\}$, the system state
is defined by four variables:

\begin{itemize}

\item $\mathbf{s}[t]$: The SIP session state,
selected from the set ( \ref{eq:sip_states})

\begin{equation}
\label{eq:sip_states}
\mathcal{S}_{\mathrm{SIP}} =
\left\{
\begin{aligned}
&\texttt{Idle},\ \texttt{Trying},\
\texttt{Provisional},\\
&\texttt{Confirmed},\
\texttt{Terminated}
\end{aligned}
\right\}.
\end{equation}

\item $\mathbf{m}[t]\in\{0,1\}$:
The media authorization state, where 1 indicates
that the media gate is open.

\item $\mathbf{c}[t]\in\{0,1\}$:
The charging record state, where 1 indicates
that the modeled charging event has occurred.

\item $\mathbf{e}[t]\in[0,T_{\max}]$:
The elapsed duration associated with media
authorization, bounded by $T_{\max}$.

\end{itemize}

The complete system state at step $t$ is represented as (Eq. \ref{eq:session_state_vector})

\begin{equation}
\label{eq:session_state_vector}
\mathbf{x}[t] =
\langle
\mathbf{s}[t],
\mathbf{m}[t],
\mathbf{c}[t],
\mathbf{e}[t]
\rangle.
\end{equation}

These variables describe the relationship between
signaling progression, media availability, and
charging behavior within the modeled network
configuration.

\subsubsection{Protocol Constraints}
\label{subsec:transition_constraints}

The model defines transition constraints between
consecutive session states.

For all $t\in\{1,\ldots,T-1\}$, the following rules
describe the relevant system behavior.

\paragraph{Media Authorization}

Entering the provisional state permits media
transmission and initializes the elapsed duration (\ref{eq:ta_entry}):

\begin{equation}
\label{eq:ta_entry}
\begin{aligned}
\big(
\mathbf{s}[t+1] &= \texttt{Provisional}
\ \land \\
\mathbf{s}[t] &\neq \texttt{Provisional}
\big)
\Rightarrow {}\\
\big(
\mathbf{m}[t+1] &= 1
\ \land\
\mathbf{e}[t+1] = 0
\big).
\end{aligned}
\end{equation}

\paragraph{Early Media Persistence}

If the session remains provisional, the media
gate remains open and the elapsed duration
increases as(\ref{eq:ta_persist}):

\begin{equation}
\label{eq:ta_persist}
\begin{aligned}
\big(
\mathbf{s}[t] &= \texttt{Provisional}
\ \land \\
\mathbf{s}[t+1] &= \texttt{Provisional}
\big)
\Rightarrow {}\\
\big(
\mathbf{m}[t+1] &= 1
\ \land\
\mathbf{e}[t+1] = \mathbf{e}[t]+\Delta t
\big).
\end{aligned}
\end{equation}

Here, $\Delta t$ represents the elapsed time between
consecutive modeled states, measured in seconds.
For the Early Media Billing Bypass scenario,
$\Delta t=15$ seconds, consistent with the
\texttt{step\_seconds} parameter in the MiniZinc
specification provided in
Appendix~\ref{app:model_billing_bypass}.

\paragraph{Session Confirmation and Charging}

The modeled charging event is associated with
a transition to the confirmed state as seen in Eq. \ref{eq:ta_charge}:

\begin{equation}
\label{eq:ta_charge}
\mathbf{s}[t+1]=\texttt{Confirmed}
\Rightarrow
\mathbf{c}[t+1]=1.
\end{equation}

This constraint represents the charging assumption
used in the evaluated scenario. It does not imply
that every IMS implementation initiates charging
exclusively after session confirmation.

\paragraph{Session Termination}

When the session reaches the terminated state,
the media gate closes as shown in Eq. \ref{eq:ta_close}:

\begin{equation}
\label{eq:ta_close}
\mathbf{s}[t+1]=\texttt{Terminated}
\Rightarrow
\mathbf{m}[t+1]=0.
\end{equation}

Together, these constraints define the relevant
state relationships for the billing bypass scenario.
The model examines whether the absence of a suitable
duration restriction can permit media authorization
to continue without the expected charging event.

\subsubsection{Security Properties and Reachability Analysis}
\label{subsec:safety_property_and_negation}

The security objective is to prevent media
transmission from remaining authorized beyond
a specified duration without the expected
charging event.

Let $T_{\mathrm{guard}}$ denote the maximum
permitted duration of uncharged early media.
The unsafe condition is defined as (\ref{eq:phi_pred})

\begin{equation}
\label{eq:phi_pred}
\begin{aligned}
\Phi(t) \triangleq {}&
\mathbf{s}[t]=\texttt{Provisional}
\ \land\\
&
\mathbf{m}[t]=1
\ \land\\
&
\mathbf{c}[t]=0
\ \land\\
&
\mathbf{e}[t]>T_{\mathrm{guard}}.
\end{aligned}
\end{equation}

The safety property requires that this
condition must not occur within the modeled
execution (\ref{eq:p_safe}):

\begin{equation}
\label{eq:p_safe}
P_{\mathrm{safe}}:
\quad
\forall t\in\{1,\ldots,T\},
\quad \neg\Phi(t).
\end{equation}

To search for a potential violation, the
safety property is negated as seen in Eq. \ref{eq:p_safe_negation}:

\begin{equation}
\label{eq:p_safe_negation}
\exists t\in\{1,\ldots,T\}:
\quad \Phi(t).
\end{equation}

A satisfying assignment for
Eq.~\eqref{eq:p_safe_negation} represents
a bounded execution in which media remains
authorized beyond the guard duration without
the modeled charging event.

Conversely, an unsatisfiable result indicates
that the specified violation is unreachable
within the selected bounds and constraints.
Such a result does not establish security
for implementations or execution conditions
outside the model.

\subsubsection{Witness Trace and Result Interpretation}
\label{subsec:witness_trace}

A satisfying assignment, when obtained,
can be represented as a state sequence (Eq. \ref{eq:witness_trace})

\begin{equation}
\label{eq:witness_trace}
\sigma =
\langle
\mathbf{x}[1],
\mathbf{x}[2],
\ldots,
\mathbf{x}[T]
\rangle.
\end{equation}

Listing~\ref{lst:trace_model1} presents the
satisfying assignment obtained from executing
the MiniZinc model. The solver identified an
unsafe state at step 7, where the session
remains provisional, media authorization
is enabled, and the elapsed duration reaches
75 seconds without the modeled charging event,
exceeding the 60-second guard threshold.

\begin{lstlisting}[
caption={MiniZinc execution output for
the Early Media Billing Bypass reachability query.},
label={lst:trace_model1}
]
EARLY MEDIA REACHABILITY QUERY
Violation step: 7
Guard duration: 60 seconds
Step duration: 15 seconds

STATE TRACE:
t=1 state=Trying media=false charged=false elapsed=0
t=2 state=Provisional media=true charged=false elapsed=0
t=3 state=Provisional media=true charged=false elapsed=15
t=4 state=Provisional media=true charged=false elapsed=30
t=5 state=Provisional media=true charged=false elapsed=45
t=6 state=Provisional media=true charged=false elapsed=60
t=7 state=Provisional media=true charged=false elapsed=75
\end{lstlisting}

The relevant sequence consists of three stages:

\begin{enumerate}

\item \textbf{Session Initiation:}
A SIP INVITE initiates the session, while the
media gate remains closed and no charging
event has occurred.

\item \textbf{Early Media Establishment:}
The session enters the provisional state.
Under the modeled media authorization policy,
the media gate opens and the elapsed duration
is initialized.

\item \textbf{Provisional State Persistence:}
The session remains provisional while media
transmission is permitted. If the elapsed
duration exceeds $T_{\mathrm{guard}}$
without the expected charging event, the
unsafe condition is reached.

\end{enumerate}

The obtained witness trace confirms that the
defined unsafe state is reachable within the
modeled configuration and execution bounds.
Its applicability to operational IMS deployments
depends on the underlying media authorization,
session timeout, and charging assumptions.

\subsection{Model 2: Unauthorized Diameter Cx Information Disclosure}

The second model examines a potential information
disclosure scenario involving an unauthorized
Diameter peer and the IMS Cx interface. The
objective is to determine whether insufficient
peer authorization enforcement permits a UAR
to be accepted and a UAA containing protected
subscriber or routing information to reach an
unauthorized requester.

The model represents peer authorization,
request acceptance, and response disclosure
as explicit conditions. It also models the
ordering of Diameter signaling events to
ensure that the request is processed before
the corresponding response is received.

The security analysis focuses on the Diameter
signaling trust boundary rather than the
timing of SIP Digest authentication. The
resulting reachability query examines whether
unauthorized information disclosure is possible
under the specified configuration assumptions.

\subsubsection{Model Representation and Assumptions}

The model represents the interaction between
an unauthorized Diameter peer, the Diameter
signaling infrastructure, and the Home
Subscriber Server (HSS).

The execution consists of five events as seen in Eq. \ref{eq:uar_event_set}:

\begin{equation}
\label{eq:uar_event_set}
\mathcal{E} =
\left\{
\begin{aligned}
&\texttt{UAR\_sent},\\
&\texttt{UAR\_forwarded},\\
&\texttt{HSS\_lookup},\\
&\texttt{UAA\_returned},\\
&\texttt{Response\_received}
\end{aligned}
\right\}.
\end{equation}

Each event is assigned a discrete timestamp
$\tau[e]\in\{1,\ldots,N\}$, where
$N=|\mathcal{E}|$.

Distinct timestamps are required for the
represented events (\ref{eq:unique_event_times}):

\begin{equation}
\label{eq:unique_event_times}
\mathrm{allDifferent}
\left(\{\tau[e]\mid e\in\mathcal{E}\}\right).
\end{equation}

The model additionally defines three Boolean
conditions: peer authorization, request
acceptance, and protected response disclosure.
These conditions are used to evaluate the
Diameter Cx security property.

\subsubsection{Protocol Constraints}

The modeled Diameter Cx exchange follows a
sequence of dependent events. A UAR is sent
and forwarded toward the HSS, which processes
the authorization request before returning a UAA.

The required event ordering is defined in Eq. \ref{eq:diameter_order}:

\begin{equation}
\label{eq:diameter_order}
\begin{aligned}
\tau_{\mathrm{UAR\_sent}}
&<
\tau_{\mathrm{UAR\_forwarded}}\\
&<
\tau_{\mathrm{HSS\_lookup}}\\
&<
\tau_{\mathrm{UAA\_returned}}\\
&<
\tau_{\mathrm{Response\_received}}.
\end{aligned}
\end{equation}

This ordering captures the causal relationships
between request generation, forwarding,
processing, and response delivery.

The model also represents the authorization
decision at the Diameter signaling boundary.
Let $A$ denote peer authorization, $E$
denote peer authorization enforcement,
and $S$ denote the availability of
protected response information.

Request acceptance $R$ is modeled as shown in Eq. \ref{eq:cx_request_acceptance}:

\begin{equation}
\label{eq:cx_request_acceptance}
R = A \lor \neg E.
\end{equation}

Protected response disclosure $D$ is modeled as shown in Eq. \ref{eq:cx_response_disclosure}:

\begin{equation}
\label{eq:cx_response_disclosure}
D = R \land S.
\end{equation}

These relationships describe the assumed
vulnerable configuration. They are abstractions
of the authorization and information-disclosure
decisions, rather than a complete implementation
of Diameter Cx processing.

\subsubsection{Security Properties and Reachability Analysis}

The security objective of this scenario is
to prevent unauthorized Diameter peers from
obtaining protected subscriber authorization
or internal IMS routing information through
the Cx interface.

Let $A$ indicate that the requesting Diameter
peer is authorized, $R$ indicate that its UAR
is accepted and processed, and $D$ indicate
that protected response information is
disclosed to the requester.

The security property is defined in Eq. \ref{eq:security_invariant}:

\begin{equation}
\label{eq:security_invariant}
P_{\mathrm{Cx}}:
\quad
\neg(\neg A \land R \land D).
\end{equation}

This property requires that an unauthorized
peer must not both obtain acceptance of its
authorization request and receive protected
response information.

To examine a potential violation, the
corresponding reachability query is defined in Eq. \ref{eq:reachability_query}:
\begin{equation}
\label{eq:reachability_query}
\exists\,\sigma:
\quad
\neg A \land R \land D.
\end{equation}

The modeled attack assumes that the adversary
can reach the Diameter signaling infrastructure.
Successful exploitation additionally requires
insufficient peer authorization or signaling
boundary enforcement and a UAA containing
information useful to the adversary.

The temporal ordering of UAR processing and UAA
delivery remains relevant to constructing a valid
protocol execution. However, the ordering of
Diameter authorization relative to SIP Digest
authentication is not itself treated as a
security violation.

A satisfying assignment to the reachability
query demonstrates that unauthorized information
disclosure is consistent with the modeled
assumptions. It does not independently establish
that a deployed IMS network permits such access.

\subsubsection{Witness Trace and Result Interpretation}

Listing~\ref{lst:trace_model2} presents the
satisfying assignment obtained from executing
the MiniZinc model. The solver confirmed that
the unauthorized disclosure condition is
reachable under the specified assumptions:
peer authorization enforcement is disabled,
the unauthorized request is accepted, and
protected response information is disclosed.
The witness trace also satisfies the required
ordering of Diameter Cx signaling events.

\begin{lstlisting}[
caption={MiniZinc execution output for
unauthorized Diameter Cx information disclosure.},
label={lst:trace_model2}
]

DIAMETER Cx DISCLOSURE QUERY
Peer authorized: false
Peer authorization enforced: false
Request accepted: true
Sensitive response available: true
Response disclosed: true

EVENT TIMESTAMPS:
UAR_sent : 1
UAR_forwarded : 2
HSS_lookup : 3
UAA_returned : 4
Response_received : 5
\end{lstlisting}

The MiniZinc witness trace in
Listing~\ref{lst:trace_model2} confirms that
the unauthorized disclosure condition is
reachable under the modeled assumptions.
With peer authorization enforcement disabled,
the unauthorized UAR is accepted, allowing
the modeled response disclosure to occur
in accordance with
Eq.~\eqref{eq:reachability_query}.
This result does not establish exploitability
in a deployed IMS network or imply that
every UAA contains protected information.

When peer authorization enforcement is enabled,
the modeled unauthorized request is rejected,
preventing the disclosure condition from being
satisfied. This illustrates the role of
Diameter peer authorization in protecting
subscriber and routing information.

\subsection{Solver Configuration and Reproducibility}

Both verification scenarios are represented
using MiniZinc constraint models. The first
examines bounded SIP session states, media
authorization, and charging conditions.
The second examines Diameter Cx peer
authorization, request acceptance, and
potential disclosure of protected response
information.

The models are intended to support reproducibility
by expressing the relevant variables, transition
rules, and security properties in an executable form.

The executable specifications are provided in
Appendix~\ref{app:minizinc_models}.
Reproduction of the analysis requires recording
the MiniZinc version, solver backend, model
parameters, and resulting solver status.

The interpretation of each result depends
on the following outcomes:

\begin{itemize}

\item \textbf{Satisfiable:}
A violating assignment exists within the
specified constraints and bounds.

\item \textbf{Unsatisfiable:}
No violating assignment exists within
the specified model and bounds.

\item \textbf{Unknown or Incomplete:}
The solver has not established either
result within the available execution
conditions.

\end{itemize}

A satisfying assignment should be accompanied
by the corresponding witness trace, including
the values of the relevant state variables
or event timestamps.

An unsatisfiable result should be reported
with the corresponding model bounds and
assumptions. Table~\ref{tab:solver_performance}
summarizes the configurations of both
MiniZinc models.

\begin{table}[htbp]
\caption{MiniZinc Model Configurations}
\label{tab:solver_performance}
\centering
\footnotesize
\renewcommand{\arraystretch}{1.15}

\resizebox{\columnwidth}{!}{%
\begin{tabular}{|p{2.7cm}|p{3.2cm}|p{3.2cm}|}
\hline
\textbf{Parameter} &
\textbf{Model 1} &
\textbf{Model 2} \\
\hline

Target protocol &
SIP early media &
Diameter Cx \\
\hline

Verification objective &
Uncharged media beyond guard duration &
Unauthorized response disclosure \\
\hline

Execution model &
Bounded session states &
Ordered Diameter events \\
\hline

Primary assumptions &
Provisional media enabled; charging on confirmation &
Peer authorization enforcement configurable \\
\hline

Default horizon &
7 state steps &
5 signaling events \\
\hline

Verification method &
Constraint satisfaction &
Constraint satisfaction \\
\hline

\end{tabular}%
}
\end{table}

Both MiniZinc models produced satisfying
assignments for their respective
security-property violation queries.
Model 1 identified an unsafe provisional
media state at step 7, while Model 2
identified an unauthorized disclosure
condition under the specified assumptions.

The experiments were executed using MiniZinc
IDE version 2.10.1 with the Gecode solver
(version 6.4). The Early Media Billing
Bypass model used a seven-step execution
horizon, a 60-second guard duration, and
a 15-second time increment. The Unauthorized
Diameter Cx Information Disclosure model
used five ordered signaling events, with
peer authorization enforcement disabled
and sensitive response information assumed
available.

The MiniZinc environment reported total
execution times of 539 ms and 341 ms for
Models 1 and 2, respectively. These durations
include execution overhead and should not
be interpreted as isolated solver computation
times or scalability measurements.

\subsection{Extending STRIDE Through MASAR}

MASAR is designed to complement and extend
STRIDE rather than replace it. STRIDE provides
the foundation for identifying and classifying
architectural security threats in Layer 1 of
the proposed methodology. MASAR builds upon
these findings by introducing additional
analytical stages that connect identified
threats to logical attack dependencies,
protocol-level security conditions, and
operational threat information.

Table~\ref{tab:comparison_strict} illustrates
how STRIDE-based threat identification is
extended through the MASAR methodology.

\begin{table}[htbp]
\caption{Extension of STRIDE-Based Threat Analysis
Through MASAR}
\label{tab:comparison_strict}
\centering
\footnotesize
\renewcommand{\arraystretch}{1.15}

\resizebox{\columnwidth}{!}{%
\begin{tabular}{|p{2.8cm}|p{3.1cm}|p{3.7cm}|}
\hline
\textbf{Analytical Stage} &
\textbf{Purpose} &
\textbf{Role in MASAR} \\
\hline

STRIDE-Based Threat 

Identification &
Identify architectural 

threats and
security concerns &
Provides the initial threat classification
in Layer 1. \\
\hline

Logical Attack Decomposition &
Represent attack prerequisites
and dependencies &
Extends identified STRIDE threats
into DAG-based attack representations
in Layer 2. \\
\hline

Protocol-Level Constraint Analysis &
Examine protocol states and
unsafe-state conditions &
Formalizes selected attack conditions
using MiniZinc constraints. \\
\hline

Operational TTP Synthesis &
Relate protocol behavior
to adversarial techniques &
Connects identified actions to
existing or candidate telecom TTPs
in Layer 3. \\
\hline

Security Control Identification &
Identify controls associated
with attack conditions &
Supports the derivation of
potential architectural and
protocol-level countermeasures. \\
\hline

\end{tabular}%
}
\end{table}

The integration of these stages enables
MASAR to connect STRIDE-based architectural
threat identification with logical attack
decomposition, protocol-state analysis,
and operational TTP synthesis. Accordingly,
MASAR should be understood as an extension
of architectural threat modeling rather
than an alternative to STRIDE.

The table describes methodological
capabilities rather than experimentally
established performance improvements.
Quantitative conclusions regarding
effectiveness, scalability, or analyst
consistency require further evaluation.

\section{Discussion: Gap Analysis and Architectural Countermeasures}
\label{sec:discussion}

The MASAR methodology aims to connect architectural security threats with specific protocol behaviors and operational security requirements. The IMS evaluation illustrates how interactions between signaling procedures, network states, and trust boundaries can create security risks that are not fully described by architectural threat classification alone.

The two examined scenarios highlight different
security concerns. Early Media Free Ride involves
the relationship between session establishment,
media authorization, and charging, while
Unauthorized Diameter Cx Information Disclosure
concerns the authorization of Diameter signaling
requests and the potential exposure of subscriber
or internal IMS routing information. Both
scenarios demonstrate the importance of examining
protocol conditions alongside architectural
security controls.

\subsection{The Signaling Plane Security Gap}

Telecommunications security frameworks provide valuable methods for identifying and classifying threats across network components. However, classifying a threat does not necessarily explain the signaling sequence or system conditions required for its occurrence.

The IMS scenarios illustrate this distinction. In the Early Media Free Ride scenario, the potential weakness arises from the relationship between provisional SIP signaling, media authorization, and charging policies. The security risk depends on whether media transmission can continue without the expected charging event or an appropriate duration restriction. Such risks are also recognized in the SIP early media security considerations described in RFC 3960 \cite{ietfEarlyMedia3960}.

Similarly, the Unauthorized Diameter Cx Information
Disclosure scenario concerns the relationship
between Diameter peer authorization and subscriber
information access. Potential disclosure occurs
when an unauthorized requesting entity receives
Diameter authorization responses without sufficient
enforcement of the relevant trust boundaries.
Depending on the returned information, this may
enable subscriber enumeration or reveal internal IMS routing characteristics.

These scenarios demonstrate why threat classification and protocol analysis should be considered together. MASAR addresses this requirement by connecting architectural threat categories with logical attack dependencies and the corresponding protocol conditions.

\subsection{The Role of Constraint Solving in Protocol Security Analysis}

Constraint solving provides a structured approach for examining relationships between protocol states, signaling events, and security properties. It allows potential attack conditions to be represented as logical constraints and evaluated within a defined model.

In MASAR, constraint satisfaction is used to examine whether selected signaling scenarios can reach states that violate specified security requirements. The resulting assignments provide information about the conditions and event sequences associated with those states.

This approach offers three main benefits:

\begin{itemize}

    \item \textbf{Structured Representation of Protocol Behavior:}
    Protocol states, events, and security conditions can be expressed as variables and logical constraints. This provides a consistent representation of the interactions examined in each attack scenario.

    \item \textbf{Connection to Logical Attack Decomposition:}
    The atomic actions and dependencies identified in the MASAR attack graphs provide a basis for defining relevant protocol conditions. Constraint models can then examine the reachability of selected unsafe states under these conditions.

    \item \textbf{Support for Reproducible Security Analysis:}
    Explicit variables, constraints, and solver configurations allow the same model to be evaluated independently. This supports the examination of security properties under clearly defined assumptions.
\end{itemize}

However, the conclusions obtained through constraint solving depend on the completeness and accuracy of the model. A satisfying assignment establishes the reachability of a state under the specified constraints, but it does not independently demonstrate that the same behavior occurs in every implementation.

Similarly, bounded verification does not establish the absence of vulnerabilities outside the modeled execution limits. The results should therefore be interpreted in relation to the selected protocol assumptions, state variables, and verification bounds.

Future extensions could incorporate additional protocol constraints, automated model generation, and optimization methods for identifying attack paths that require fewer actions or expose more significant security risks.

\subsection{Architectural Countermeasures}

The MASAR analysis provides a basis for identifying security controls associated with the conditions represented in each attack scenario. Rather than relying only on broad threat categories, the proposed countermeasures focus on the signaling procedures and trust boundaries relevant to the modeled weaknesses.

\subsubsection{Mitigating Early Media Exploitation}

The Early Media Free Ride scenario highlights the need to coordinate SIP signaling states with media authorization and charging policies.

Three security measures are relevant to this scenario:

\begin{itemize}

    \item \textbf{Media Duration Controls:}
    Network operators should enforce configurable limits on early media sessions. When the permitted duration is exceeded, the network should apply the appropriate charging policy or terminate unauthorized media transmission.

    \item \textbf{State Dependent Media Authorization:}
    Media gateways and session border functions should enforce media authorization policies according to session state and service requirements. Early media transmission should remain subject to the configured restrictions even when final SIP session confirmation has not occurred.

    \item \textbf{Charging and Signaling Coordination:}
    Charging mechanisms should account for media usage according to the applicable service policy. This includes ensuring that prolonged provisional sessions do not bypass the charging controls intended for the service.
\end{itemize}

These measures are consistent with the early media risks and possible remedies described in RFC 3960 \cite{ietfEarlyMedia3960}. The SIP P-Early-Media mechanism also provides a means to communicate early media authorization policies within applicable trust domains \cite{ietfEarlyMedia5009}.

The appropriate configuration depends on the operator's service model, charging architecture, and permitted early media behavior. Consequently, restrictions should be selected according to the deployment requirements rather than applying a universal prohibition on bidirectional early media.

\subsubsection{Securing the Diameter Cx Interface}

The Unauthorized Diameter Cx Information Disclosure
scenario highlights the importance of controlling
access to subscriber authorization procedures and
protecting sensitive information returned through
Diameter signaling.

The following measures address the relevant trust boundaries:

\begin{itemize}

    \item \textbf{Diameter Peer Authentication and Authorization:}
    Diameter signaling infrastructure should accept authorization requests only from authenticated and authorized peers. The requesting entity's identity and permitted signaling functions should be validated before requests are forwarded toward subscriber information services.

    \item \textbf{Signaling Boundary Enforcement:}
    Diameter routing and edge security functions should restrict access to internal authorization services. Network segmentation, access control policies, and appropriate signaling protection mechanisms can reduce the ability of unauthorized entities to submit subscriber queries.

    \item \textbf{Response Information Control:}
    Authorization responses should be handled according to the requesting peer's permissions and the applicable protocol requirements. Access to subscriber status and internal routing information should be restricted to entities authorized to receive it.

    \item \textbf{Monitoring of Authorization Requests:}
    Security monitoring can identify unusual patterns of subscriber authorization queries, including repeated requests involving multiple subscriber identities or unexpected Diameter peers.
\end{itemize}

These measures address the potential disclosure of subscriber and routing information through insufficient signaling authorization. Their implementation should follow the relevant Diameter and IMS security requirements, including the procedures defined in 3GPP TS 29.229 \cite{3gpp29229}.

The proposed controls focus on the Diameter signaling trust boundary rather than assuming that SIP subscriber authentication must precede every legitimate Cx authorization procedure.

\subsection{Operational Contribution to MITRE FiGHT}

The final MASAR layer connects the protocol actions identified through logical decomposition with operational threat information. This process supports the classification of attack behavior and the identification of candidate technique descriptions for telecom security knowledge bases.

Table~\ref{tab:proposed_ttps} summarizes the two evaluated scenarios, their associated STRIDE categories, and the candidate operational techniques derived from the analysis.

\begin{table}[htbp]
\centering
\caption{Candidate Operational Techniques Derived from MASAR}
\label{tab:proposed_ttps}
\footnotesize
\renewcommand{\arraystretch}{1.15}

\resizebox{\columnwidth}{!}{%
\begin{tabular}{|p{2.4cm}|p{2.4cm}|p{4.1cm}|}
\hline
\textbf{Attack Scenario} &
\textbf{STRIDE Category} &
\textbf{Candidate Technique Description} \\
\hline

Early Media Free Ride &
Tampering &
Manipulation of provisional SIP session behavior to maintain media transmission without the expected charging event. \\
\hline

Unauthorized Diameter Cx Information Disclosure &
Information Disclosure &
Use of unauthorized Diameter authorization queries
to infer subscriber information or internal IMS
routing characteristics. \\

\hline

\end{tabular}%
}
\end{table}

To characterize the two candidate techniques identified through
MASAR, three technical dimensions are considered:
(1) Protocol Layer and Interface,
(2) Prerequisite State Context, and
(3) Action Granularity.
Table~\ref{tab:ttp_differentiation} summarizes these
characteristics and highlights the technical distinctions
of the proposed techniques.
The candidate identifiers are provisional suggestions
and have not been assigned or approved by MITRE.
Their relationship to existing MITRE FiGHT techniques
remains subject to independent review and classification.

\begin{table*}[htbp]
\caption{Technical Characterization of Proposed
MITRE FiGHT Candidate Techniques}
\label{tab:ttp_differentiation}
\centering
\footnotesize
\renewcommand{\arraystretch}{1.2}
\begin{tabular}{|p{1.7cm}|p{3.1cm}|p{3.1cm}|p{6.9cm}|}
\hline
\textbf{Provisional ID} &
\textbf{Candidate Technique} &
\textbf{Target Protocol} &
\textbf{Technical Characteristics} \\
\hline
\textbf{FGT-C-1001} &
Provisional-State Early-Media Exploitation &
SIP and RTP/SRTP &
Exploitation of provisional SIP session states
and early-media authorization where session
duration or charging controls are insufficient. \\
\hline
\textbf{FGT-C-1002} &
Unauthorized Diameter Cx Subscriber and
Topology Information Disclosure &
Diameter Cx &
Unauthorized UAR signaling accepted across an
insufficiently enforced Diameter trust boundary,
potentially disclosing subscriber authorization
outcomes or internal IMS routing information. \\
\hline
\end{tabular}
\end{table*}

These candidate descriptions identify the protocol behavior and security objectives relevant to each scenario. Their relationship to existing MITRE FiGHT techniques should be established through comparison with the current knowledge base and related security research.

The operational value of the MASAR synthesis process lies in linking architectural threats with identifiable protocol actions and potential detection requirements. This relationship can support security monitoring, incident investigation, and the development of controls targeting specific attack conditions.

The proposed technique descriptions should therefore be treated as candidates for further assessment rather than confirmed additions to MITRE FiGHT.

\subsection{Limitations and Future Research Directions}

The MASAR evaluation focuses on two IMS signaling
scenarios and uses constraint-based models to examine
selected security properties. While these scenarios
illustrate the proposed methodology, several
limitations affect the interpretation and
generalizability of the findings.

First, the current implementation follows an
analyst-guided process. STRIDE threat mapping,
attack DAG construction, operational TTP mapping,
and MiniZinc constraint model development are
performed manually. Although this approach supports
traceable security analysis, it requires domain
expertise and may introduce analyst-dependent
variations. Future work will investigate automating
these transformations and assessing their consistency
across different analysts and network configurations.

Second, the constraint models represent selected
protocol behaviors rather than complete IMS
implementations. Their conclusions are limited
by the defined states, transitions, assumptions,
and execution bounds. Validation in a representative
IMS emulation environment is therefore needed to
determine whether the modeled attack conditions
correspond to observable behavior in practical
deployments.

Third, empirical scalability and performance
measurements are not yet available. Although the
number of primary state variables in the Early
Media Billing Bypass model increases linearly
with the execution horizon, this does not imply
linear solver runtime or memory consumption.
Future experiments will evaluate the finalized
MiniZinc models across different execution
bounds and configurations, recording solver
runtime, memory usage, and search statistics.

Fourth, operational TTP mapping requires further
validation against existing telecom threat
knowledge bases and documented attack behavior.
Establishing consistent criteria for technique
classification and novelty assessment remains
an important direction for future research.

To contextualize the formal verification approach,
Table~\ref{tab:formal_tools_comp} summarizes the
different analytical purposes and representations
of established formal methods and the
constraint-based modeling approach adopted
by MASAR.

\begin{table}[htbp]
\caption{Qualitative Comparison of Formal Verification
Approaches in Telecom Security}
\label{tab:formal_tools_comp}
\centering
\footnotesize
\renewcommand{\arraystretch}{1.15}
\setlength{\tabcolsep}{3pt}

\resizebox{\columnwidth}{!}{%
\begin{tabular}{|p{2.2cm}|p{2.5cm}|p{2.5cm}|p{2.5cm}|}
\hline
\textbf{Approach} &
\textbf{Primary Focus} &
\textbf{State Representation} &
\textbf{Analysis Output} \\
\hline

SPIN / LTL &
Software protocol model checking &
Automata / Promela &
State-space exploration and counterexamples \\
\hline

UPPAAL &
Real-time system verification &
Timed automata &
Timed-state reachability analysis \\
\hline

TAMARIN &
Cryptographic protocol verification &
Multiset rewriting &
Security proofs and counterexample traces \\
\hline

\textbf{MASAR (MiniZinc)} &
\textbf{Telecom cross-layer security analysis} &
\textbf{Finite-domain CSP constraints} &
\textbf{Bounded unsafe-state reachability queries} \\
\hline

\end{tabular}%
}
\end{table}

Unlike the general-purpose verification approaches
summarized in Table~\ref{tab:formal_tools_comp},
MASAR uses MiniZinc constraint modeling to represent
selected protocol states and security properties
derived from architectural threat analysis and
logical attack decomposition. This supports bounded
reasoning about potential unsafe states, although
comparative performance and verification effectiveness
have not yet been experimentally established.

Finally, extending the evaluation to additional
telecom architectures, protocol implementations,
and attack scenarios will help assess the
generalizability, reproducibility, and practical
applicability of MASAR.

\section{Conclusion and Future Work}
\label{sec:conclusion}

This paper presented MASAR, a structured
threat modeling methodology that connects
architectural security risks with protocol-level
adversarial behavior in telecommunications
networks. MASAR integrates strategic threat
mapping using STRIDE, logical attack path
decomposition using directed acyclic graphs,
and operational TTP synthesis informed by
telecom security knowledge bases, including
MITRE FiGHT.

The methodology was applied to a 5G-ready
IMS architecture through two signaling
attack scenarios: Provisional-State
Early-Media Exploitation using SIP, and
Unauthorized Diameter Cx Subscriber and
Topology Information Disclosure.

The analysis illustrates how architectural
weaknesses can be translated into explicit
attack conditions and protocol-level
security properties. MiniZinc constraint
models support bounded reachability analysis
under defined assumptions. The resulting
candidate technique descriptions provide
potential contributions to telecom threat
knowledge bases, subject to independent
review and classification.

The current transformation process is
analyst-guided and manually performed.
The verification results are conditional
on the modeled architecture, protocol
behavior, and configuration assumptions;
they do not independently establish
exploitability in operational IMS networks.

Future work will focus on automating the transformation of architectural threat models into attack graphs and executable constraint models, reducing the need for manual analysis. Further evaluation across different telecom architectures and representative network implementations will also be considered to assess the framework's scalability, reproducibility, and effectiveness. In addition, integrating MASAR into continuous security assessment processes could support the identification of emerging risks as network configurations and services evolve.

\appendix
\section{Executable MiniZinc Models}
\label{app:minizinc_models}

This appendix provides the MiniZinc specifications for the two constraint-based analyses described in Section~\ref{sec:formal_verification}.

The models represent the selected protocol states,
authorization conditions, event ordering constraints,
and reachability queries used to examine the Early
Media Free Ride and Unauthorized Diameter Cx
Information Disclosure scenarios.

The specifications were executed using
MiniZinc IDE version 2.10.1 with the
Gecode solver (version 6.4).
Both models use finite domains and bounded execution traces. Their results must therefore be interpreted according to the stated assumptions, parameters, and execution bounds.

\lstset{
  language=MiniZinc,
  basicstyle=\ttfamily\tiny, 
  numbers=left,
  numberstyle=\tiny\color{gray},
  stepnumber=1,
  numbersep=3pt,
  backgroundcolor=\color{gray!5},
  showspaces=false,
  showstringspaces=false,
  showtabs=false,
  frame=single,
  tabsize=1,
  captionpos=b,
  breaklines=true, 
  breakatwhitespace=false, 
  xleftmargin=10pt,
  xrightmargin=2pt
}

\subsection{Model 1: Early Media Billing Bypass}
\label{app:model_billing_bypass}

The first model examines whether an IMS session can remain in a provisional state with an open media gate and no charging record beyond a specified duration.

\begin{lstlisting}[
caption={MiniZinc model for early media persistence and the billing bypass reachability query.},
label={lst:early_media_model}
]
include "globals.mzn";

% Model parameters
int: T = 7;
int: T_guard = 60;
int: T_max = 120;
int: step_seconds = 15;

% SIP states
enum SipState = {
    Idle, Trying, Provisional, Confirmed, Terminated
};

% State variables
array[1..T] of var SipState: sip_state;
array[1..T] of var bool: media_open;
array[1..T] of var bool: charge_record;
array[1..T] of var 0..T_max: elapsed;

% Initial conditions
constraint sip_state[1] = Trying;
constraint media_open[1] = false;
constraint charge_record[1] = false;
constraint elapsed[1] = 0;

% Valid session transitions
constraint forall(t in 1..T-1)(
    (sip_state[t] = Trying ->
        (sip_state[t+1] = Trying \/
         sip_state[t+1] = Provisional \/
         sip_state[t+1] = Terminated))
    /\
    (sip_state[t] = Provisional ->
        (sip_state[t+1] = Provisional \/
         sip_state[t+1] = Confirmed \/
         sip_state[t+1] = Terminated))
    /\
    (sip_state[t] = Confirmed ->
        (sip_state[t+1] = Confirmed \/
         sip_state[t+1] = Terminated))
    /\
    (sip_state[t] = Terminated ->
        sip_state[t+1] = Terminated)
);

% Media gate policy for modeled configuration
constraint forall(t in 1..T)(
    media_open[t] =
        (sip_state[t] = Provisional \/
         sip_state[t] = Confirmed)
);

% Charging event persists after confirmation
constraint forall(t in 1..T-1)(
    charge_record[t+1] =
        (charge_record[t] \/
         sip_state[t+1] = Confirmed)
);

% Elapsed time rules
constraint forall(t in 1..T-1)(
    if sip_state[t+1] = Provisional then
        if sip_state[t] = Provisional then
            elapsed[t+1] = elapsed[t] + step_seconds
        else
            elapsed[t+1] = 0
        endif
    else
        elapsed[t+1] = 0
    endif
);

% Negated safety property
var 1..T: viol_t;

constraint
    sip_state[viol_t] = Provisional
    /\ media_open[viol_t] = true
    /\ charge_record[viol_t] = false
    /\ elapsed[viol_t] > T_guard;


solve satisfy;

output
    ["EARLY MEDIA REACHABILITY QUERY\n",
     "Violation step: ", show(viol_t), "\n",
     "Guard duration: ", show(T_guard), " seconds\n",
     "Step duration: ", show(step_seconds), " seconds\n\n",
     "STATE TRACE:\n"]
    ++
    [
        "t=" ++ show(t)
        ++ " state=" ++ show(sip_state[t])
        ++ " media=" ++ show(media_open[t])
        ++ " charged=" ++ show(charge_record[t])
        ++ " elapsed=" ++ show(elapsed[t])
        ++ "\n"
        | t in 1..T
    ];

\end{lstlisting}

\subsection{Model 2: Unauthorized Diameter Cx Information Disclosure}

\label{app:model_uar_leak}

The second model examines whether insufficient
Diameter Cx peer authorization enforcement
allows an unauthorized requester to obtain
protected subscriber or IMS routing information.
The analysis represents authorization decisions,
response disclosure conditions, and the ordering
of UAR/UAA signaling events.

\begin{lstlisting}[
caption={MiniZinc model for unauthorized Diameter Cx information disclosure.},
label={lst:uar_model}
]
include "globals.mzn";

% Configuration assumptions
bool: peer_authorized = false;
bool: enforce_peer_authorization = false;
bool: sensitive_response_available = true;

% Authorization and disclosure decisions
var bool: request_accepted;
var bool: response_disclosed;

% Model the authorization boundary
constraint
    request_accepted =
        (peer_authorized \/
         not enforce_peer_authorization);

% A sensitive response is disclosed only when
% the request is accepted and relevant data exists
constraint
    response_disclosed =
        (request_accepted /\
         sensitive_response_available);

% Successful UAR/UAA event ordering
enum Ev = {
    UAR_sent,
    UAR_forwarded,
    HSS_lookup,
    UAA_returned,
    Response_received
};

int: N = card(Ev);
array[Ev] of var 1..N: event_time;

constraint
    all_different([event_time[e] | e in Ev]);

constraint
    event_time[UAR_sent] <
    event_time[UAR_forwarded];

constraint
    event_time[UAR_forwarded] <
    event_time[HSS_lookup];

constraint
    event_time[HSS_lookup] <
    event_time[UAA_returned];

constraint
    event_time[UAA_returned] <
    event_time[Response_received];

% Negated security property:
% unauthorized peer receives protected data
constraint
    not peer_authorized /\
    request_accepted /\
    response_disclosed;

solve satisfy;

output
    ["DIAMETER Cx DISCLOSURE QUERY\n",
     "Peer authorized: ",
         show(peer_authorized), "\n",
     "Peer authorization enforced: ",
         show(enforce_peer_authorization), "\n",
     "Request accepted: ",
         show(request_accepted), "\n",
     "Sensitive response available: ",
         show(sensitive_response_available), "\n",
     "Response disclosed: ",
         show(response_disclosed), "\n\n",
     "EVENT TIMESTAMPS:\n"]
    ++
    [
        show(e) ++ " : " ++
        show(event_time[e]) ++ "\n"
        | e in Ev
    ];
\end{lstlisting}

\bibliographystyle{IEEEtran}
\bibliography{references}

@techreport{3gpp23501,
  author       = {{3GPP}},
  title        = {System architecture for the {5G} System ({5GS})},
  institution  = {3rd Generation Partnership Project (3GPP)},
  number       = {TS 23.501},
  year         = {2024},
  note         = {Version 18.5.0, Release 18}
}

@article{gong2025survey,
  author       = {Gong, Mengyi and Wei, Ziling and Chen, Shuhui and Yu, Wanrong and Wang, Fei},
  title        = {A survey of existing attacks on {5G SA}},
  journal      = {Computer Networks},
  volume       = {270},
  pages        = {111545},
  year         = {2025},
  doi          = {10.1016/j.comnet.2025.111545}
}

@article{tariq2025survey,
  author       = {Tariq, Saman and Rodriguez Luna, Eva and Masip Bruin, Xavier and Diaz, Rodrigo and Martrat, Josep and Trakadas, Panagiotis},
  title        = {A survey on {5G} private and {B5G} network threats and safeguarding {AI}-based security mechanisms through the layered analysis},
  journal      = {Computer Networks},
  volume       = {271},
  pages        = {111594},
  year         = {2025},
  doi          = {10.1016/j.comnet.2025.111594}
}

@techreport{3gpp33501,
  author       = {{3GPP}},
  title        = {Security architecture and procedures for {5G} System ({5GS})},
  institution  = {3rd Generation Partnership Project (3GPP)},
  number       = {TS 33.501},
  year         = {2025},
  note         = {Version 18.8.0, Release 18}
}

@article{breeden2026survey,
  author       = {Breeden, John and Khan, Mohammad S. and Silvestri, Simone and Bajracharya, Biju and Scott, Chandler},
  title        = {Mitigating slice vulnerabilities in {5G} core networks: A comprehensive survey},
  journal      = {Ad Hoc Networks},
  volume       = {183},
  pages        = {104118},
  year         = {2026},
  doi          = {10.1016/j.adhoc.2025.104118}
}

@book{shostack2014threat,
  author       = {Shostack, Adam},
  title        = {Threat Modeling: Designing for Security},
  publisher    = {John Wiley \& Sons},
  year         = {2014},
  isbn         = {978-1-118-80999-0}
}

@book{ucedavelez2015risk,
  author       = {UcedaVelez, Marco M. and Morana, Marco M.},
  title        = {Risk Centric Threat Modeling: Process for Attack Simulation and Threat Analysis (PASTA)},
  publisher    = {John Wiley \& Sons},
  year         = {2015},
  isbn         = {978-1-118-98835-0}
}

@techreport{mitrefight2023,
  author       = {Radle, Andrew J. and Arnoth, Eric I. and Stephenson, Amir and Vanderveen, Michaela and Ahmed, Muddasar S. and Dey, Surajit and Garcia Gonzalez, Adrian},
  title        = {{MITRE FiGHT: High-Level Overview (Five-G Hierarchy of Threats)}},
  institution  = {The MITRE Corporation},
  number       = {MP230545},
  year         = {2023},
  month        = {August},
  url          = {https://fight.mitre.org/FiGHT_High-Level_Overview_PRS-23-2698.pdf}
}

@article{unger2026attackgraphs,
  author       = {Unger, Simon and Arzoglou, Ektor and Heinrich, Markus and Scheuermann, Dirk and Katzenbeisser, Stefan},
  title        = {Cybersecurity risk assessment in {OT} systems using attack graphs},
  journal      = {International Journal of Information Security},
  volume       = {25},
  pages        = {34},
  year         = {2026},
  doi          = {10.1007/s10207-025-01198-7}
}

@article{rao2023bhadra,
  author       = {Rao, Siddharth Prakash and Chen, Hsin Yi and Aura, Tuomas},
  title        = {Threat modeling framework for mobile communication systems},
  journal      = {Computers \& Security},
  volume       = {125},
  pages        = {103047},
  year         = {2023},
  issn         = {0167-4048},
  doi          = {10.1016/j.cose.2022.103047}
}

@techreport{gsmaFS31,
  author       = {{GSMA}},
  title        = {{FS.31 Baseline Security Controls}},
  institution  = {GSM Association},
  type         = {Permanent Reference Document},
  year         = {2023},
  note         = {Version 3.0},
  url          = {https://www.gsma.com/solutions-and-impact/technologies/security/wp-content/uploads/2024/06/FS.31-Baseline-Security-Controls-v3.0.pdf}
}

@article{palma2025progressive,
  author       = {Palma, Alessandro and Cicimurri, Claudio and Angelini, Marco},
  title        = {Progressive attack graph: a technique for scalable and adaptive attack graph generation},
  journal      = {International Journal of Information Security},
  volume       = {24},
  pages        = {212},
  year         = {2025},
  doi          = {10.1007/s10207-025-01125-w}
}

@article{sainthilaire2025attackdefense,
  author       = {Saint-Hilaire, K{\'e}ren and Neal, Christopher and Cuppens, Fr{\'e}d{\'e}ric and Boulahia Cuppens, Nora and Bassi, Francesca and Hadji, Makhlouf},
  title        = {A real-time automated attack-defense graph generation approach},
  journal      = {Journal of Information Security and Applications},
  volume       = {94},
  pages        = {104266},
  year         = {2025},
  doi          = {10.1016/j.jisa.2025.104266}
}

@article{layered2022electronics,
  author       = {Farooqui, Muhammad Najmul Islam and Arshad, Junaid and Khan, Muhammad Mubashir},
  title        = {A Layered Approach to Threat Modeling for {5G}-Based Systems},
  journal      = {Electronics},
  volume       = {11},
  number       = {12},
  pages        = {1819},
  year         = {2022},
  doi          = {10.3390/electronics11121819}
}

@article{kordy2014dag,
  author       = {Kordy, Barbara and Pi{\`e}tre-Cambac{\'e}d{\`e}s, Ludovic and Schweitzer, Patrick},
  title        = {{DAG}-Based Attack and Defense Modeling: Don't Miss the Forest for the Attack Trees},
  journal      = {Computer Science Review},
  volume       = {13--14},
  pages        = {1--38},
  year         = {2014},
  doi          = {10.1016/j.cosrev.2014.07.001}
}

@article{widel2019beyond,
  author       = {Wide{\l}, Wojciech and Audinot, Maxime and Fila, Barbara and Pinchinat, Sophie},
  title        = {Beyond 2014: Formal Methods for Attack Tree-Based Security Modeling},
  journal      = {ACM Computing Surveys},
  volume       = {52},
  number       = {4},
  pages        = {75:1--75:36},
  year         = {2019},
  doi          = {10.1145/3331524}
}

@article{konsta2024survey,
  author       = {Konsta, Alyzia Maria and Lluch-Lafuente, Alberto and Spiga, Beatrice and Dragoni, Nicola},
  title        = {Survey: Automatic Generation of Attack Trees and Attack Graphs},
  journal      = {Computers \& Security},
  volume       = {137},
  pages        = {103602},
  year         = {2024},
  doi          = {10.1016/j.cose.2023.103602}
}

@article{zhang2025attackgraphs,
  author       = {Zhang, Luyao and Tang, Gaigai and He, Xin and Qi, Kaiyuan and Su, Guangfeng and Zhang, Huiyun},
  title        = {Automatic generation of industrial internet attack graphs with graph neural networks and Bayesian models},
  journal      = {Computer Networks},
  volume       = {272},
  pages        = {111736},
  year         = {2025},
  doi          = {10.1016/j.comnet.2025.111736}
}

@article{jelba2026zerotrust,
  author       = {Jelba, J. and Sajan, R. Isaac},
  title        = {Zero-Trust Oriented Threat Detection and Orchestration Across Data and Control Planes in {5G} Cloud Networks},
  journal      = {International Journal of Communication Systems},
  volume       = {39},
  number       = {12},
  pages        = {e70541},
  year         = {2026},
  doi          = {10.1002/dac.70541}
}

@inproceedings{rupprecht2020revolte,
  author       = {Rupprecht, David and Kohls, Katharina and P{\"o}pper, Christina and Holz, Thorsten},
  title        = {Call Me Maybe: Eavesdropping Encrypted {LTE} Calls With {ReVoLTE}},
  booktitle    = {29th USENIX Security Symposium},
  pages        = {73--88},
  year         = {2020},
  publisher    = {USENIX Association}
}

@inproceedings{lee2022vwanalyzer,
  author       = {Lee, Hyunwoo and Jeong, Jaejong and Kim, Jinhong and Kim, Myungbae and Kim, Hyoungshick and Kang, Brent Byunghoon},
  title        = {{VWAnalyzer}: A Systematic Security Analysis Framework for the Voice over {WiFi} Protocol},
  booktitle    = {Proceedings of the 2022 ACM Asia Conference on Computer and Communications Security},
  pages        = {999--1013},
  year         = {2022},
  doi          = {10.1145/3488932.3517415}
}

@inproceedings{li2015volte,
  author       = {Li, Chi-Yu and Tu, Guan-Hua and Peng, Chunyi and Yuan, Zengwen and Li, Yuanjie and Lu, Songwu and Wang, Xinbing},
  title        = {Insecurity of Voice Solution {VoLTE} in {LTE} Mobile Networks},
  booktitle    = {Proc. 22nd ACM CCS},
  pages        = {316--327},
  year         = {2015},
  doi          = {10.1145/2810103.2813618}
}

@inproceedings{lu2020ghost,
  author       = {Lu, Yu-Han and Li, Chi-Yu and Li, Yao-Yu and Hsiao, Sandy H.-Y. and Xie, Tian and Tu, Guan-Hua and Chen, Wei-Xun},
  title        = {Ghost Calls from Operational {4G} Call Systems: {IMS} Vulnerability, Call {DoS} Attack, and Countermeasure},
  booktitle    = {Proc. 26th MobiCom},
  pages        = {96--109},
  year         = {2020},
  doi          = {10.1145/3372224.3380885}
}

@article{catta2025formal,
  author       = {Catta, Davide and Leneutre, Jean and Malvone, Vadim and Murano, Aniello},
  title        = {A Formal Approach to Attack Graphs},
  journal      = {Annals of Mathematics and Artificial Intelligence},
  volume       = {93},
  number       = {4},
  pages        = {589--610},
  year         = {2025},
  doi          = {10.1007/s10472-024-09959-1}
}

@article{coral2025,
  author       = {Tayouri, David and Sgan Cohen, Omri and Maimon, Inbar and Mimran, Dudu and Elovici, Yuval and Shabtai, Asaf},
  title        = {{CORAL}: Container Online Risk Assessment with Logical Attack Graphs},
  journal      = {Computers \& Security},
  volume       = {150},
  pages        = {104296},
  year         = {2025},
  doi          = {10.1016/j.cose.2024.104296}
}

@book{clrs2022introduction,
  author       = {Cormen, Thomas H. and Leiserson, Charles E. and Rivest, Ronald L. and Stein, Clifford},
  title        = {Introduction to Algorithms},
  publisher    = {MIT Press},
  edition      = {4th},
  year         = {2022},
  isbn         = {978-0262046305}
}

@techreport{ietfEarlyMedia3960,
  author       = {Camarillo, Gonzalo and Schulzrinne, Henning},
  title        = {Early Media and Ringing Tone Generation in the Session Initiation Protocol ({SIP})},
  institution  = {RFC Editor},
  type         = {RFC},
  number       = {3960},
  year         = {2004},
  month        = {December},
  doi          = {10.17487/RFC3960},
  url          = {https://www.rfc-editor.org/rfc/rfc3960}
}

@techreport{ietfSIP3261,
  author       = {Rosenberg, J. and Schulzrinne, H. and Camarillo, G. and Johnston, A. and Peterson, J. and Sparks, R. and Handley, M. and Schooler, E.},
  title        = {{SIP: Session Initiation Protocol}},
  institution  = {RFC Editor},
  number       = {3261},
  type         = {RFC},
  year         = {2002},
  publisher    = {IETF}
}

@techreport{3gpp29229,
  author       = {{3GPP}},
  title        = {Cx and Dx interfaces based on the Diameter protocol; Protocol details},
  institution  = {3rd Generation Partnership Project},
  type         = {Technical Specification},
  number       = {TS 29.229},
  year         = {2026},
  note         = {Version 19.1.0, Release 19},
  url          = {https://portal.3gpp.org/desktopmodules/Specifications/SpecificationDetails.aspx?specificationId=1682}
}

@techreport{ietfDiameter6733,
  author       = {Fajardo, V. and Arkko, J. and Loughney, J. and Zorn, G.},
  title        = {{Diameter Base Protocol}},
  institution  = {RFC Editor},
  number       = {6733},
  type         = {RFC},
  year         = {2012},
  publisher    = {IETF}
}

@misc{minizinc2024,
  author       = {{MiniZinc Project}},
  title        = {{The MiniZinc Handbook}, Version 2.9},
  year         = {2025},
  url          = {https://docs.minizinc.org/en/2.9.4/},
  note         = {Accessed: October 2026}
}

@techreport{ietfEarlyMedia5009,
  author       = {Ejzak, Richard P.},
  title        = {Private Header ({P-Header}) Extension to the Session Initiation Protocol ({SIP}) for Authorization of Early Media},
  institution  = {RFC Editor},
  type         = {RFC},
  number       = {5009},
  year         = {2007},
  month        = {September},
  doi          = {10.17487/RFC5009},
  url          = {https://www.rfc-editor.org/rfc/rfc5009}
}

\end{document}